\documentclass[10pt]{article}

\usepackage{amsmath}
\usepackage{amssymb}

\usepackage{subfigure}
\usepackage{graphicx}

\usepackage{cite}
\usepackage{color}

\usepackage[labelfont=bf,labelsep=period,justification=raggedright]{caption}

\date{}

\usepackage{subfigure}

\usepackage[utf8]{inputenc}

\begin{document}

\begin{flushleft}
{\Large
\textbf{Epidemics scenarios in the ``Romantic network''}
}


Alexsandro M. Carvalho$^{1}$,
Sebastián Gonçalves$^{1}$
\\
\bf{1}Instituto de Física, Universidade Federal do Rio Grande do Sul,
Caixa Postal 15051, CEP 90501-970, Porto Alegre, RS, Brazil
\\
$\ast$ E-mail: sgonc@if.ufrgs.br
\end{flushleft}

\section*{Abstract}

\textbf{Background:} The structure of sexual contacts, its contacts
network and its temporal interactions, play an important role in the
spread of sexually transmitted infections. Unfortunately, that kind of
data is very hard to obtain. One of the few exceptions is the
``Romantic network'' which is a complete structure of a real sexual
network of a high school.  In terms of topology, unlike other sexual
networks classified as scale-free network, it does not correspond to
any model network, as the authors of the work concluded based in many
network measures.  Regarding the temporal structure, several studies
indicate that relationship timing can have effects on diffusion
through networks, as relationship order determines transmission
routes.

\noindent\textbf{Methodology:} With the aim to check if the particular
structure, static and dynamic, of the Romantic network is determinant
for the propagation of an STI in it, we perform simulations in two
scenarios: the static network where all contacts are available and the
dynamic case where contacts evolve in time.  In the static case, we
compare the epidemic results in the Romantic network with some
paradigmatic topologies. We further study the behavior of the epidemic
on the Romantic network in response to three variants. First, we
investigate the effect of any individual, belonging to the network,
having a contact with an external infected subject. Second, we explore
the influence of the degree of the initial infected. Third, we analyze
the effect of the variability of contacts per unit time. Finally, we
consider the dynamics of formation of pairs in the Romantic network
and we study the propagation of the diseases in this dynamic scenario.

\noindent\textbf{Conclusions:} Our results suggest that the scale-free
network displays indeed a substantial difference with the Romantic
network. On the other side, while this real network can not be labeled
as a Watts-Strogatz network, it is, regarding the propagation of an
STI, very close to one with high disorder.  Additionally, we find in
relation to the three epidemic variations: $(i)$ the effect that any
individual contacting an external infected subject is to make the
network closer to a full connected one, $(ii)$ a higher contact degree
of the initial infected accelerates outbreaks (earlier peak time and
higher epidemic probability) and $(iii)$ the epidemic impact is
proportional to the numbers of contacts per unit time. Finally, our
simulations confirm that relationship timing affects, but strongly
lowering, the final outbreak size. Besides, shows a clear correlation
between the average degree and the outbreak size over time.


\section*{Introduction}
The progress in medical and biological sciences by the end of sec XIX
eventually led to the identification of the organisms responsible for
the diseases, making possible the systematization of the mathematical
description of epidemics.  While first attempt to mathematical model
epidemics is situated probably at 1760 with Daniel Bernoulli and his
study of vaccination on the spread of smallpox~\cite{Bernoulli}, was
Hamer who introduced the idea of mass action~\cite{Hamer} to model
epidemics.  Then Kermack and McKendrick used that idea to evolve into
the present form of the susceptible-infective-removed (SIR)
model~\cite{Kermack1991}, and its variants, so ubiquitous in the
nowadays theoretical description of epidemics~\cite{Bailey,Murray}.
As a mass action or mean field model, the SIR model (and its SIS,
SIRS, etc, variants) assumes that all classes are superimposed in
space\footnote{Even in the more elaborated case of space distributed
classes (Fisher's equations) it is locally assumed the superposition
of them.}.  However, in the last decade, it became clear that the
transmission of a disease can be influenced by the contact
network\cite{Colizza2009, Colizza2011}.  Such network depends on the
structure of interactions among the individuals, which in turn may
depend on the peculiar disease.  Several articles have addressed the
disease propagation problem on networks using paradigmatic examples
like small-world~\cite{Kuperman2001} or scale-free
networks~\cite{Albert2002}.  The high heterogeneity of some extreme
topologies lead to important consequences as the persistence of
epidemics for any value of the spreading rate in an infinite
scale-free network\footnote{The reason comes from the non epidemic
threshold of an infinite-variance degree distribution of such infinite
network, however human networks dubiously present such a extreme
behavior; for more details see
Ref~\cite{Jones2003}.}~\cite{Pastor2001,May2001}.  General results
regarding the threshold, size and distribution of epidemic bursts were
given for small-world networks in one and two
dimensions~\cite{Newman2002b}.

While there have been recent efforts to incorporate human behavior into
disease models (see ref~\cite{Funk2010} for a review), to implement
disease-spreading agent based models, including attributes at real
human populations sizes, it is yet a very difficult
task~\cite{Eubank2004}.  Before that level of sophistication became
eventually a not so hard enterprise, the study of the spreading of
hypothetical diseases in real networks can be of great help. Until
recently, the complete description of a real network ---at least one
which could be related with a specific disease--- was something that
seemed not feasible. Consequently, the only information at hand was
statistical data like the degree distribution of the network
associated with the disease. That is the case of the Sweden
study~\cite{Liljeros2001}, in which a questionnaire addressing sexual
behavior was completed by 4781 subjects.  Unfortunately, that kind of
questionnaire not only is subjected to bias\footnote{For our purposes,
the most relevant questions are about the number o partners in the
last year and in lifetime, the last one is particularly strong
biased.}~\cite{Foxman2006}, but does not allow us to build the real
network behind the individuals, because the subjects represent a small
sample of a larger population, the links of which are not known.

However, since some years ago, the picture of a small, but real,
sexual and romantic network is available. The ``Romantic Network'', as
called by the authors~\cite{Bearman2004}, is one of a few examples of
a complete network\footnote{There are other two networks more recently
available and made from real data: the Likoma
network~\cite{Helleringer2007} and the Internet sex network of
Brazil~\cite{Rocha2010}. However, from the figures of the first one it
is not possible to code the network as we did with the romantic
network. The second one was obtained indirectly from posts to a forum
and restricted to sex commerce contacts. For a broad picture of the
available human networks and how the data were obtained see
Ref~\cite{Doherty2011}.}, so simulation of epidemics on it can be very
important, even considering its restricted validity.

Moreover, information on the dynamic formation of links is available
for the Romantic network~\cite{Moody-web}, which makes this particular
example more interesting.  If we take the static network and run a
simulation of an epidemic, with some values of infectivity and
infection time, at the end we can have let say 10\% of the subjects
infected. Now, if we take into account the dynamics of formation of
links, what will be the final result?  In other words, how different
will be the final outcome from the static scenario? There is no answer
to that question without doing a simulation in the dynamic network and
that is one of the principal aims of the present contribution.

Thus, in the present contribution we implement and analyze computer
simulations of the evolution of an hypothetical sexually transmitted
infection (STI) in the Romantic network.  From such simulations it is
possible to evaluate the risk and impact of a potential epidemic in
this and similar networks.  The analysis is based on the specific
structure of that network and it is compared to several well known
model networks, which are potentially representative of human
interactions. 

Once the network is established and known, one is faced with different
schematic ways of conducting the disease spreading
simulations. Consider an infected node with some other susceptible
nodes connected to it. The different ways in which the contagion can go
from this node to its neighbors can be schematically divided in three
cases: at each time step (which can represent a day for example) it
can be through only one link, through some of them simultaneously, or
through all of them in the same period. Among the three possibilities
and for matters of simplicity we chose the first one (one contact per
time step) for all the simulations of the Results section, leaving one
specific sub-section to discuss the other possible choices.

Different from the usual approach of mass action models and static
networks, several recent
articles~\cite{Morris1997,Moody2002,Volz2007,Rissau2010,Rocha2011}
have addressed the problem of disease propagation from the perspective
of dynamic formation of links.  As was emphasized in those
contribution we believe that the timing present in all social networks
is a key factor regarding the flow of information (disease, rumors,
etc) that goes through them.  Therefore one section of our
contribution is devoted to this subject.

\section*{Materials and Methods}

\subsection*{Empirical network}
In 2004 the first picture of an almost complete sexual network of a
social group was published~\cite{Bearman2004}.  This network, called
by the authors ``Romantic and Sexual Network'' emerged from data taken
from the National Longitudinal Study of Adolescent Health, a
longitudinal study of students in grades 7-12 in US. Among the 140
schools included in the study, 14 had saturated
field-settings\footnote{In ``saturated'' cases {\bf all} students of
the school took the in-home interviews, which ask to nominate their
partners, so complete sexual and/or romantic networks could therefore
be constructed with these data.}.  From those, one of the two largest
ones ($N \approx 1000$) was used to depict the aforementioned
network. The data was obtained after home interviews covering a period
of 18 month, where students were asked to identify their sexual or
romantic partners from a list of the student attending their school.
``Jefferson High School'' is the fake name of the only public high
school in ``Jefferson City'', a non-identified mid-sized mid-western
town of US, with a homogeneous (all-white, mostly working class
students) and isolated community.  Participated in the in-school
survey 90\% of the school roster and a total of 873 completed the
questionnaires. After data was collected a network of 573 nodes
emerged, from which probably the most remarkable feature is a
ring-like component involving 288 students. The 288 nodes component
emerged as a result of superimposing all contacts among the students
during the 18 month period, see Fig. 1B.

However, taking into account the temporal sequence, James Moody
published a dynamic version of this network in his
site~\cite{Moody-web}. In order to see the evolution of an
hypothetical infection both in the static and dynamic network, the
animated figure cited above was broken into its $45$ frames, each one
corresponding to a period of 12 days, which for simplicity we assume
evenly distributed in time.  The nodes and links between them were
identified and converted to input data for the simulation of the
epidemic models.  Each frame taken individually looks very different
from the superposition of all of them (the giant static component).
To give an idea of how it looks like, we show in Fig.~\ref{frames} two
frames at times $t_1=10$ and $t_2=34$, which represent times of low
and high activity in the network, respectively. Therefore, it could not
have been guessed before the picture was finished, nor by the
researchers, neither by the participants.  Besides, we manually mapped
the graphical representation of the giant component (static and
dynamic versions) and checked with Pajek~\cite{Batagelj1998} that our
transcription gave the same graphical representation than the original
one.

\subsection*{Artificial networks}
For the artificial networks we use the Watts-Strogatz
(W-S)~\cite{Watts1998} and Krapivsky-Redner (K-R)~\cite{Krapivsky2001}
models, which are good candidates for representing human
relations. They cover many scenarios from ordered to disordered, and
from homogeneous to heterogeneous, respectively, by means of one
control parameter. Besides, they allow an easy control of the average
degree.

\begin{figure}[ht!] \centerline{
\subfigure{\includegraphics[width=3.5cm]{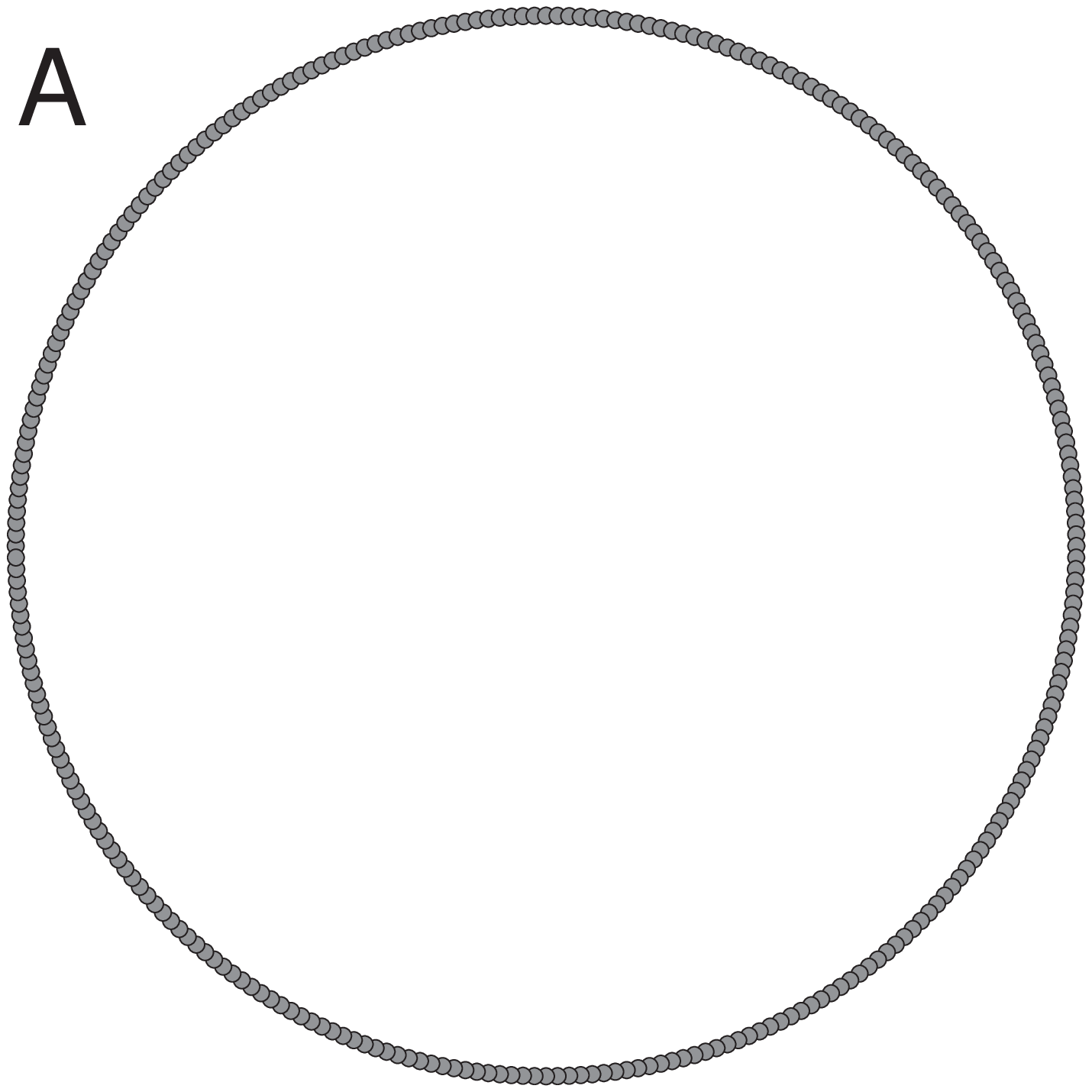}\label{fig:swp0}} \hfil
\subfigure{\includegraphics[width=3.5cm]{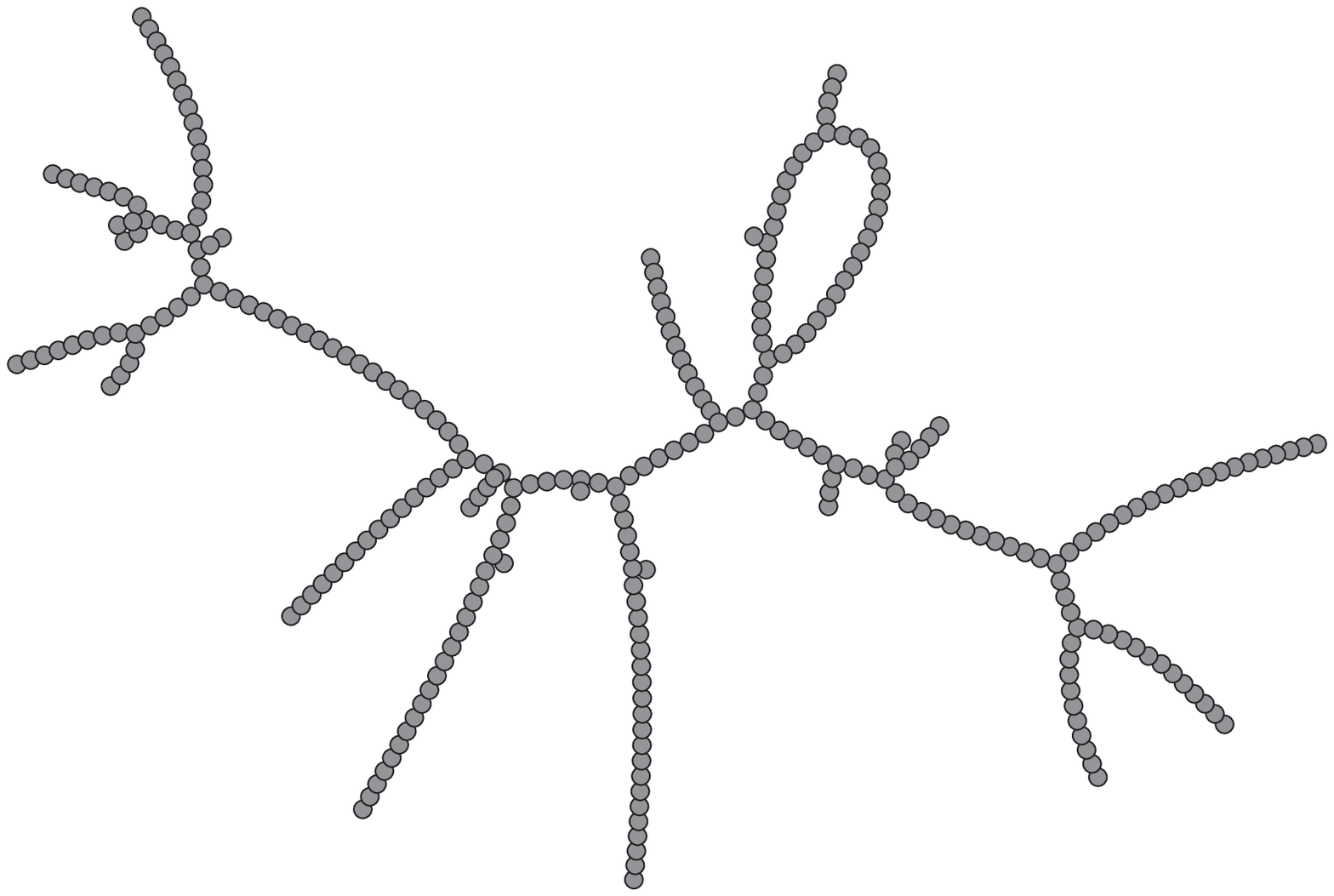}\label{fig:swp0.1}}\hfil
\subfigure{\includegraphics[width=3.5cm]{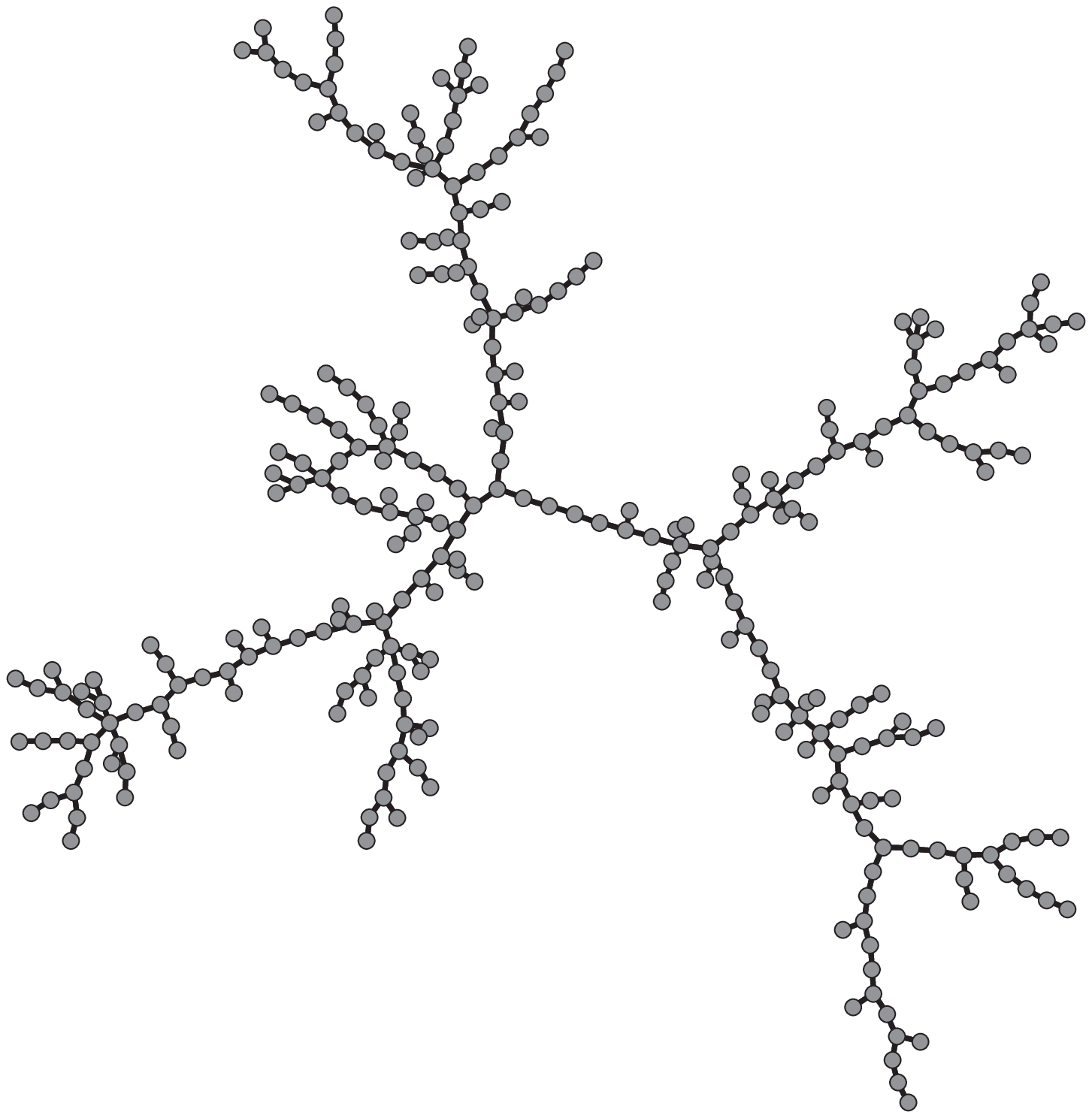}\label{fig:swp1}}
}
  \centerline{
    \subfigure{\includegraphics[width=3.5cm]{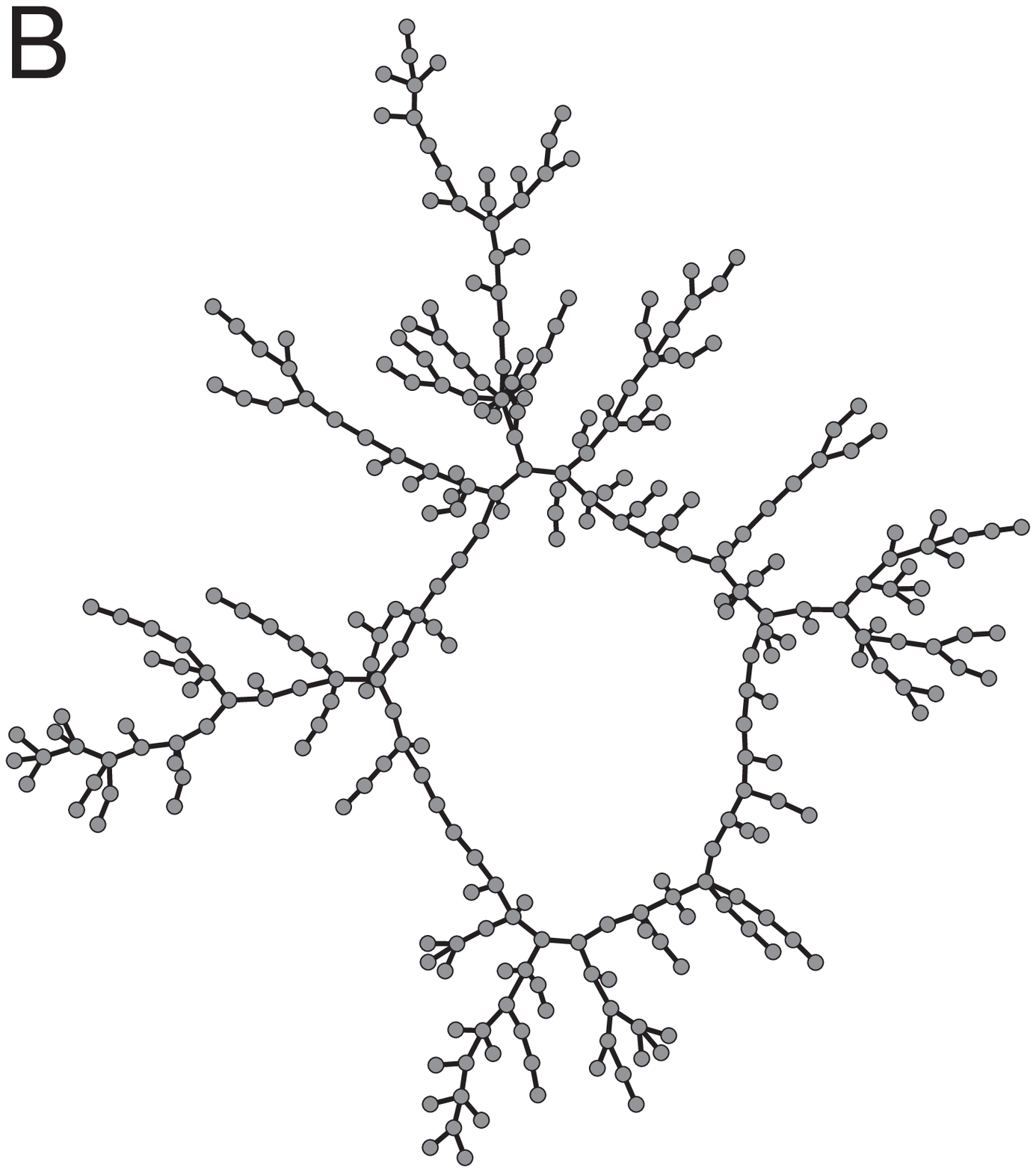}\label{fig:real}}}
  \centerline{ \subfigure{\includegraphics[width=3.5cm]{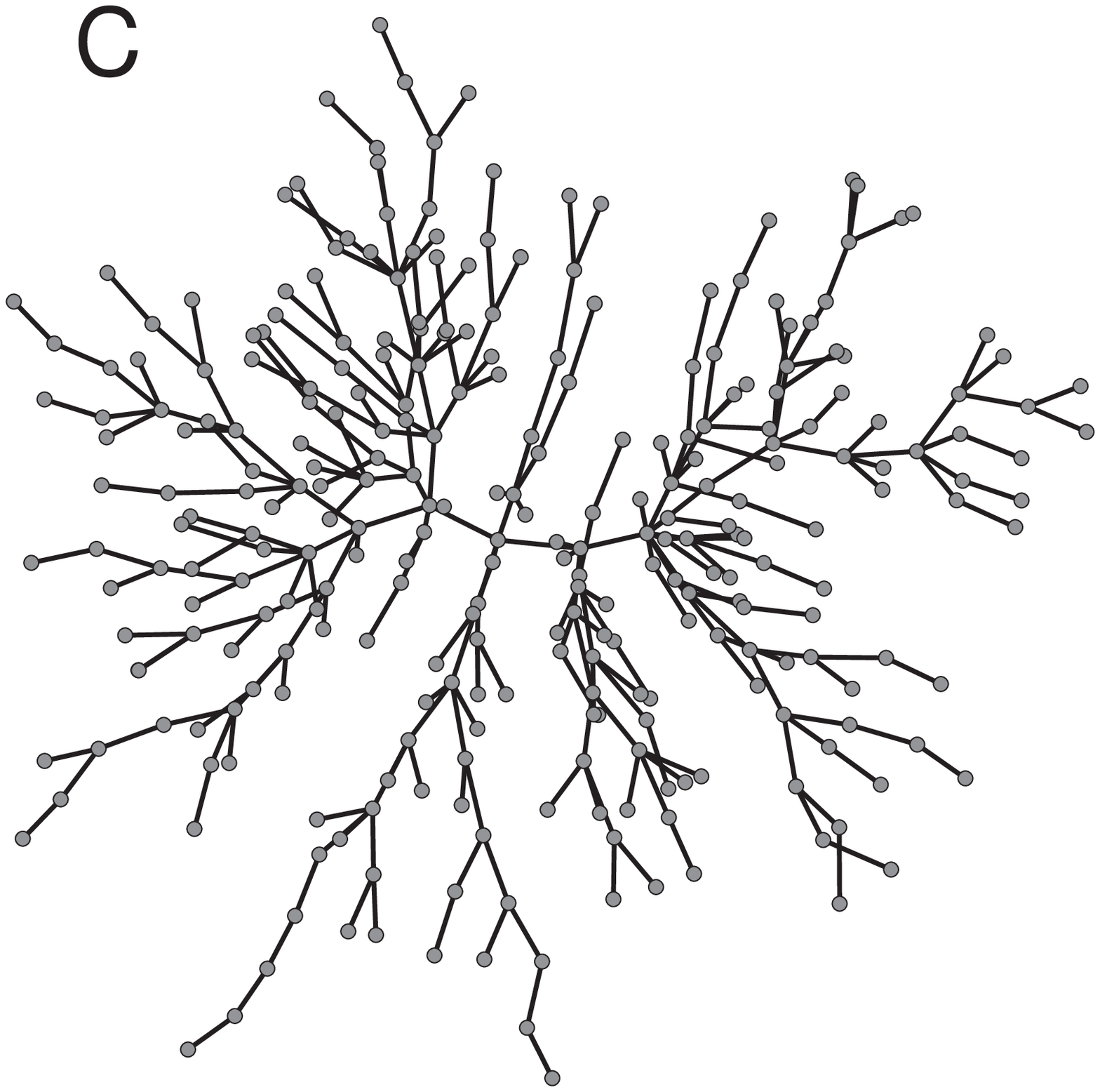}\label{fig:kr0.1}}
    \hfil \subfigure{\includegraphics[width=3.5cm]{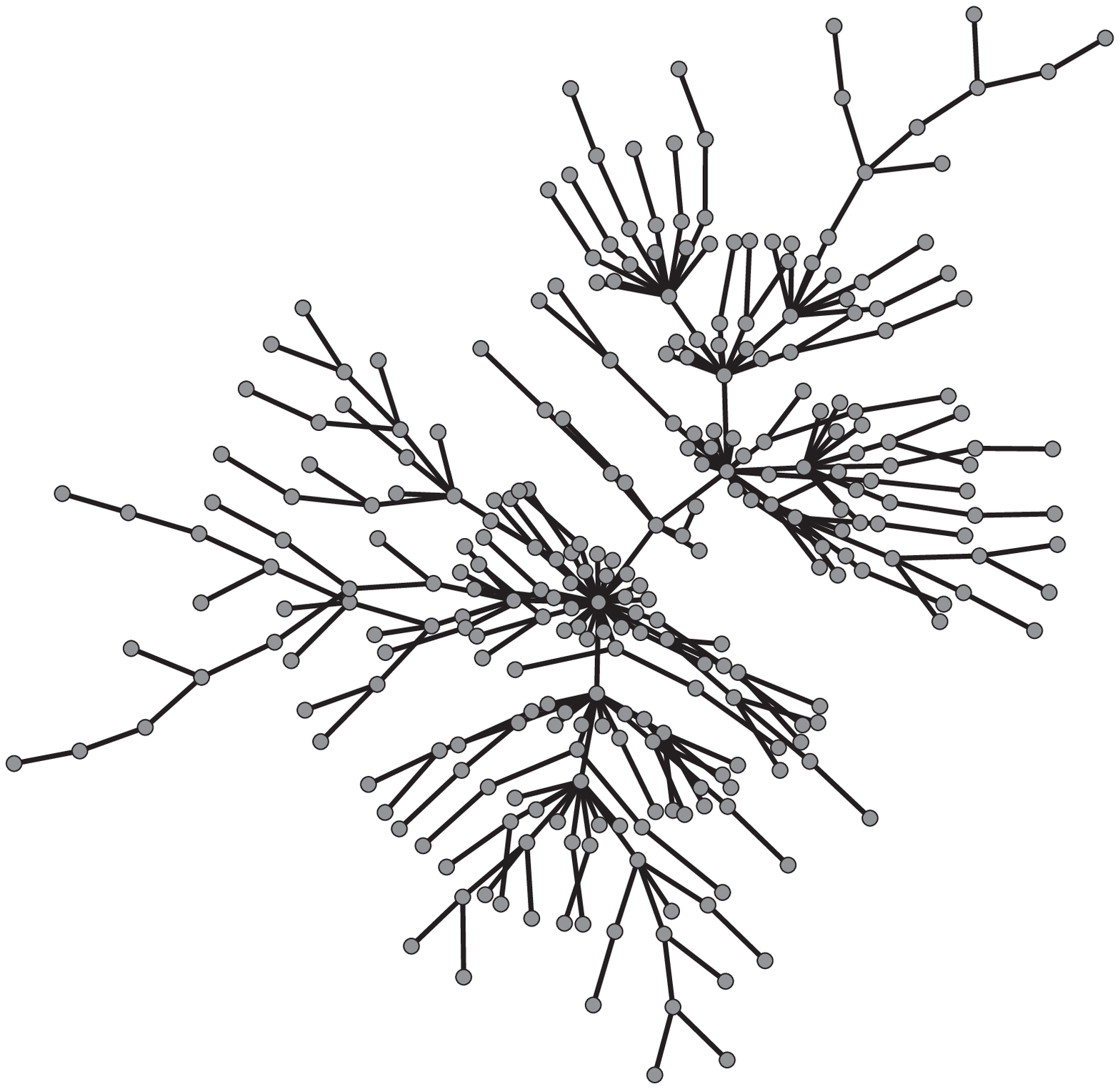}\label{fig:kr0.5}}\hfil
    \subfigure{\includegraphics[width=3.5cm]{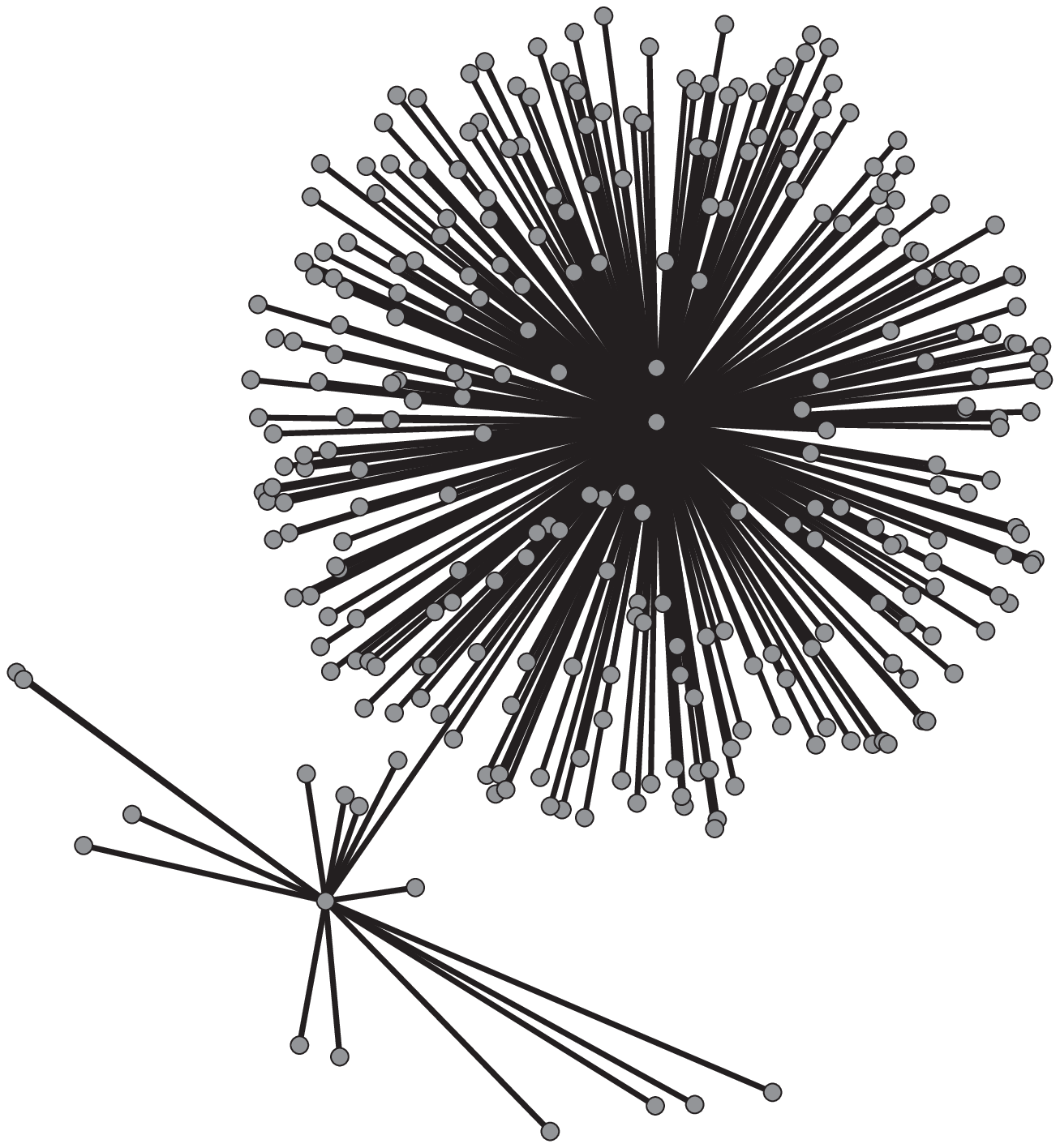}\label{fig:kr1}}
    }
\caption{{\bf Networks realizations used in this work:} (A) Watts-Strogatz:
from left to right, $p=0$ (regular), $p=0.1$, and $p=1$ (random); (B)
Romantic; (C) Krapvisky-Redner: from left to right, $r=0.1$, $r=0.5$,
and $r=1$.  All of them have the same number of vertices, $N=288$, and
the same average degree, $\left< k \right>=2$. Figures generated with
PAJEK~\cite{Batagelj1998}.}
\label{fig:networks}
\end{figure}

\begin{figure}[htb!]
\centerline{
\subfigure{\includegraphics[width=6.5cm]{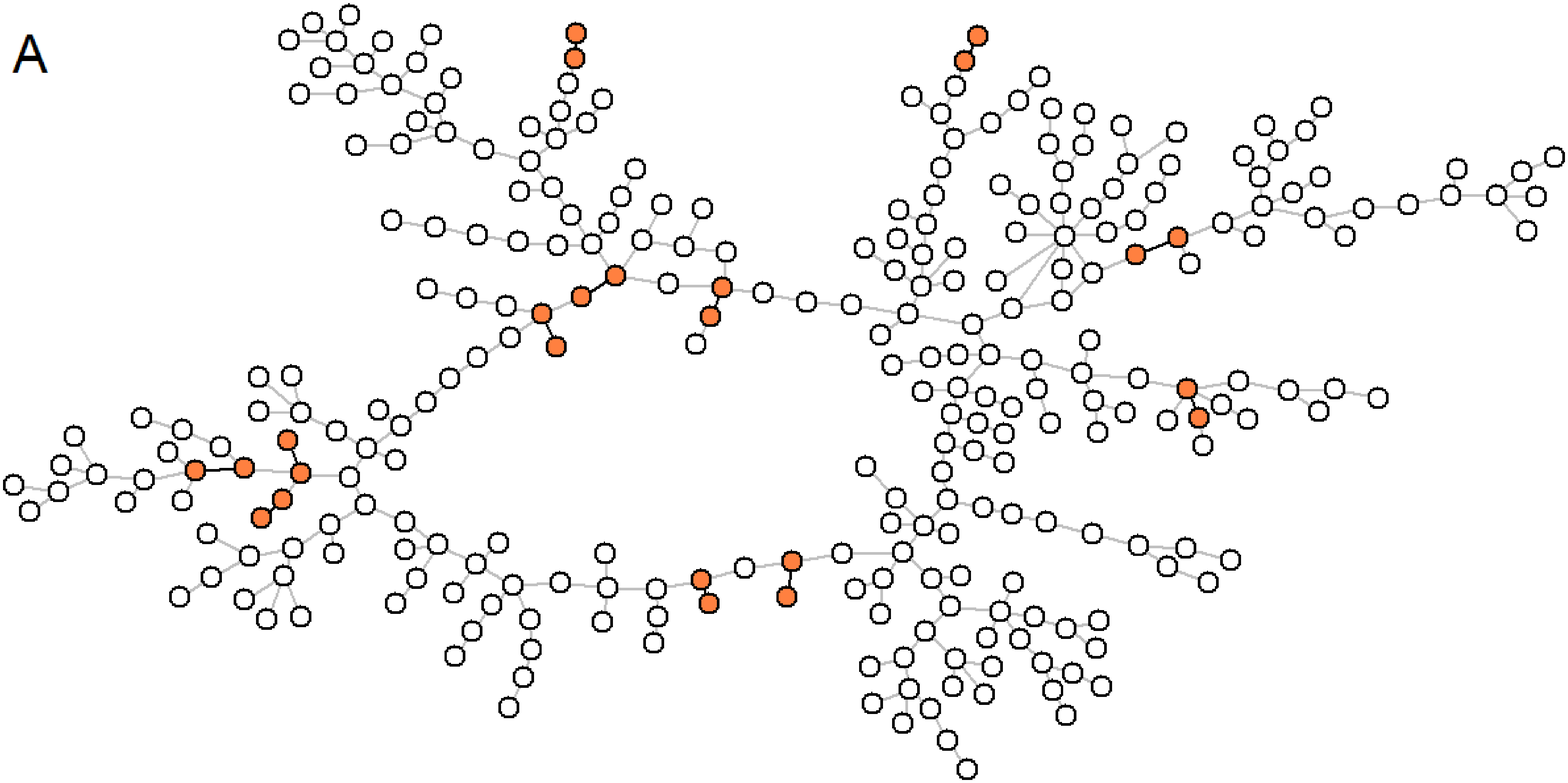}
    } \hfil
\subfigure{\includegraphics[width=6.5cm]{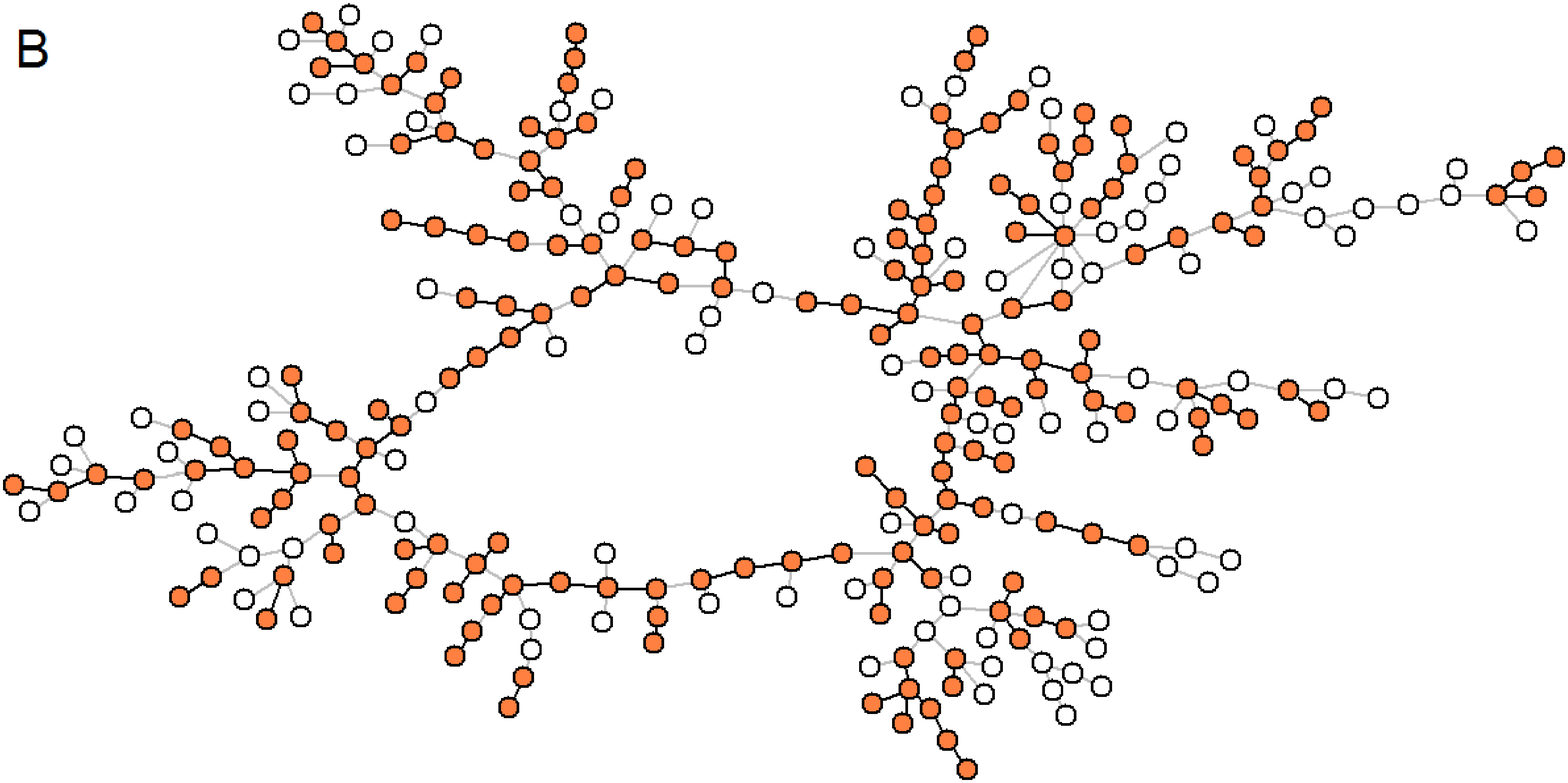}
    }}
\caption{{\bf Snapshots of the network taken from frames at times (A)
$t_1=10$ and (B) $t_2=34$.} The hatched vertices correspond to the
interacting individuals. Figures generated with PAJEK~\cite{Batagelj1998}.
}\label{frames}
\end{figure}

The W-S networks are implemented starting from a ring of 288 nodes
linked up to first neighbors. We let the disorder parameter $p$ to
have several values running from 0 (the original ordered ring) to 1 (a
complete random network), avoiding the rupture of the component, see
Fig.~\ref{fig:networks}A.  So all degrees of disorder are probed
besides the normally accepted small-world values ($p \approx 0.01$).
For the heterogeneous K-R networks we use the redirection procedure
(see Fig.~\ref{fig:networks}C): at every time step, a new node is
added to the system and an earlier node $i$ is selected uniformly, as
a possible target for attachment. With probability $1-r$ a directed
edge from the new node to $i$ is created; with probability $r$ the
edge is redirected to the ancestor node $j$ of node $i$.  Starting
from two connected nodes we let the network evolve until it reaches a
288 nodes component. This model leads eventually to a power-law degree
distribution ---when it grows to large sizes--- with degree exponent
$\gamma=1+\frac{1}{r}$, which can be tuned to any value equal or
larger than 2.

The real network is bipartite (145 males and 143 females), but for all
the epidemic dynamics in the real and artificial networks we consider
all nodes of the same kind, because we assume them as non directional
networks, without distinction between links. Apart from the size, the
most relevant feature for the implementation of model networks that
have to mimic the largest connected component of the Romantic network,
is the average degree $\left < k\right>$. Its value is $\left <
k\right> = 2$, so all artificial networks were build having this same
value.

\subsection*{Simulation of epidemics}
\label{computacional}
The epidemics in the networks are studied by computational simulations
of the SIR model. For each specific network topology and realization
---in which each vertex represents a person, with links representing
the sexual relation---, one individual is randomly chosen to be the
first infected, leaving the rest in the susceptible state.  That is
the initial state i.e., one node in the I state and the other 287
nodes in the S state. The simulation proceeds as follow: at each time
step (which could represent one day or one week), each infected person
chose a partner randomly from the list of subjects connected with
him/her. If the chosen node is in infected or removed state, nothing
happens, but if is in susceptible state it becomes infected with
probability $\beta$~\footnote{There are three random processes in the
simulations which are implemented by the use of a pseudo-random number
generator in the computer (we use Marsaglia random
generator~\cite{Marsaglia}): the selection of the first infected among
the N (288) nodes, the selection of one of the partners of each
infected node at each time step, and the stochastic process of
infection with probability $\beta$. The two first ones are like
throwing a dice, while the last one is implemented in a Monte Carlo
(MC) procedure: by generating a random number uniformly in the [0,1)
interval, and comparing with $\beta$, if the number is smaller (or
equal) than $\beta$, the infection occurs, otherwise not. Besides, a
MC procedure is used previously to generate the W-S and K-R
networks.}.  Such procedure mimics a sexual contact of two persons in
which one of them has an STI.  The simulation step is completed after
all nodes have contacted one of its partners, so we assume that the
sexual activity is homogeneous among the population and independent of
the number of partners. We will discuss other options later.  Once
infected, the subject remains so for a fixed period $\tau$ --being
potentially contagious to its list of contacts as the dynamics
proceeds. After that period, the infected individual is removed (or
becomes permanent immune). The process is repeated until no infected
individual remains in the population.  For a given situation, the
disease parameters $\beta$ and $\tau$ are fixed and have the same
value for all subjects.

The time step, as we mentioned before, can represent one day or one
week or another reasonable time period.  If it is one day, that means
that every subject is having one intimate or sexual contact per day,
which is probably an overestimation.  However, the simulation step
could be assigned to a week which can underestimate the danger of the
epidemic. Anyway, that frequency related to sexual activity is
absorbed in the $\beta \tau$ number, and a broad range of them will be
tested in this presentation.

For the static scenario, the Romantic and artificial networks are
characterized by the following quantities relevant for the SIR model:
the final size of the epidemic, $R(\infty)$, the maximum value of
infected subjects or epidemic peak, $I_{max}$, the time to arrive to
that peak, $t_p$, and the epidemic probability. All of them were
obtained computing those realizations that ended up with more than 5\%
of infected subjects. It is indeed the ratio between the number of
those realizations and the total what we use to define the probability
to observe an epidemic. The number of different realization for each
kind of network (and for each value of $\beta \tau$) is two thousand,
which corresponds to ten different graphs for each kind of network and
two hundred different random seeds of the random number generator
(i.e.  different initial infected subject) used to simulate the
infection.

Other possibilities in relation to the Romantic network are explored:
external field, degree of the initial infected, and rules of
interaction.

By external field, $B$, we mean the probability that a subject within
the static empirical network interacts with an outsider, who is
assumed to be infected.  Therefore, the chosen one, selected at random
from the entire population, is infected with probability $\beta$.
Obviously, $B=0$ is equivalent to the normal dynamics without external
excursions, while $B=1$ means that at each time step someone certainly
encounters an infected outsider.  $B \beta$ is then the likelihood of
an spontaneous infection in the network.

Then, we probe the influence of the initial infected.  In particular,
we want to see such influence regarding the degree of the node,
i.e. the number of connections of the initial infected subject. The
rationale for this is that someone with many links should have more
potential of infection than someone with only one link.  We think this
is a very important question to elucidate before going to epidemics in
real situations.  In order to test that, we define the initial
infected individual according with its degree, from the minimum value
$k_i=1$ to the maximum $k_i=9$, comparing between them and with the
random election of the initial infected.  When $k_i$ has its maximum
value there is only one option for the initial infected, because only
one subject in the static empirical network has such degree. Then, as
the degree of the initial infected goes down, i.e. exploring lower
values of $k_i$, we have more options for the initial infected (there
are more individuals with lower values of connectivity). In those
cases we average over several initial choices of subjects. Besides, in
all cases, we average over many initial random seeds, so even in the
one possibility initial case ($k_i=9$) it represents the average of
different possible ``histories'', starting from that situation.

In real social dynamics the frequency of interactions is not uniformly
distributed, i.e. one person could visit one or many of its contacts
during a fixed period of time. In order to capture this feature in the
epidemic models we face different possibilities when an infected
subject interacts with its contacts; we call them rules of
interaction. One: from all its contact, only one is chosen randomly at
each time step. Fraction: a subset of its links is selected at random
to contact this set at the same time step. All: all connected vertices
are contacted at the same time step.

The dynamic network brings a new scenario.  The inclusion of the
dynamic formation of links modulates one of the basic properties of
the network, the degree. In order to characterize the evolution of the
dynamic network we measure (using the snapshots of the animated gif)
the degree of each node over time and we use this data to obtain the
dynamic degree distribution, $P_{t}(k)$.  This distribution represents
the probability that a randomly chosen node, at a given time $t$, has
degree $k$. From this quantity, we follow the changes of the average
degree over time, $\left < k\right>_{t}$. Since at different times,
many sub-graph are disconnected from the giant component, another
relevant measure is the size of the largest connected component,
$G$. Lastly, we characterized the spread of the epidemic in this
scenario. For this, we measure the fraction of infected nodes $I$.

\section*{Results}

\subsection*{Static Scenario}
\label{staticnetworks}

\subsubsection*{Romantic network vs artificial networks}
\begin{figure}[htb!]
\centerline{
\subfigure{\includegraphics[width=7.5cm]{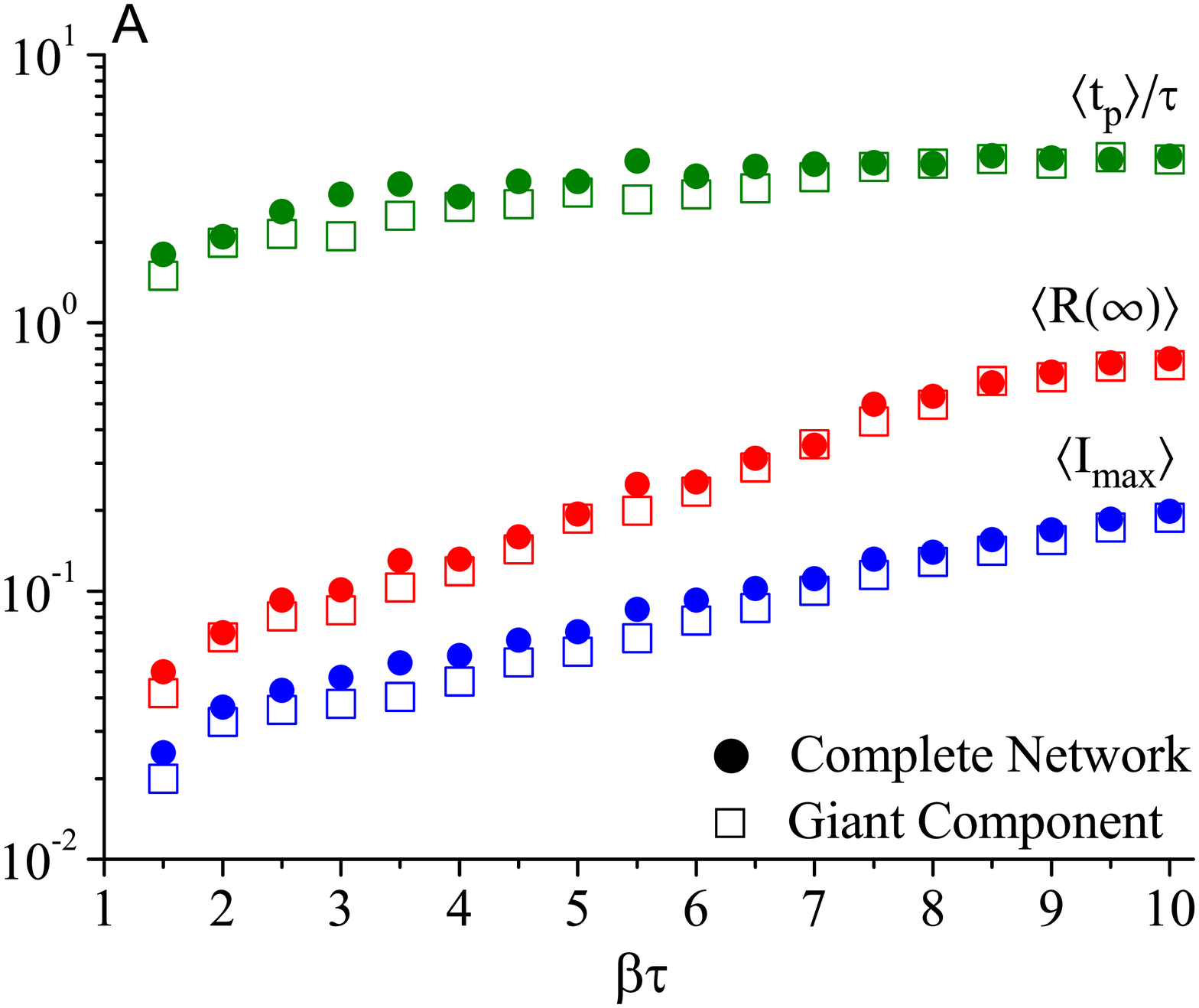}
   } \hfil
\subfigure{\includegraphics[width=7.5cm]{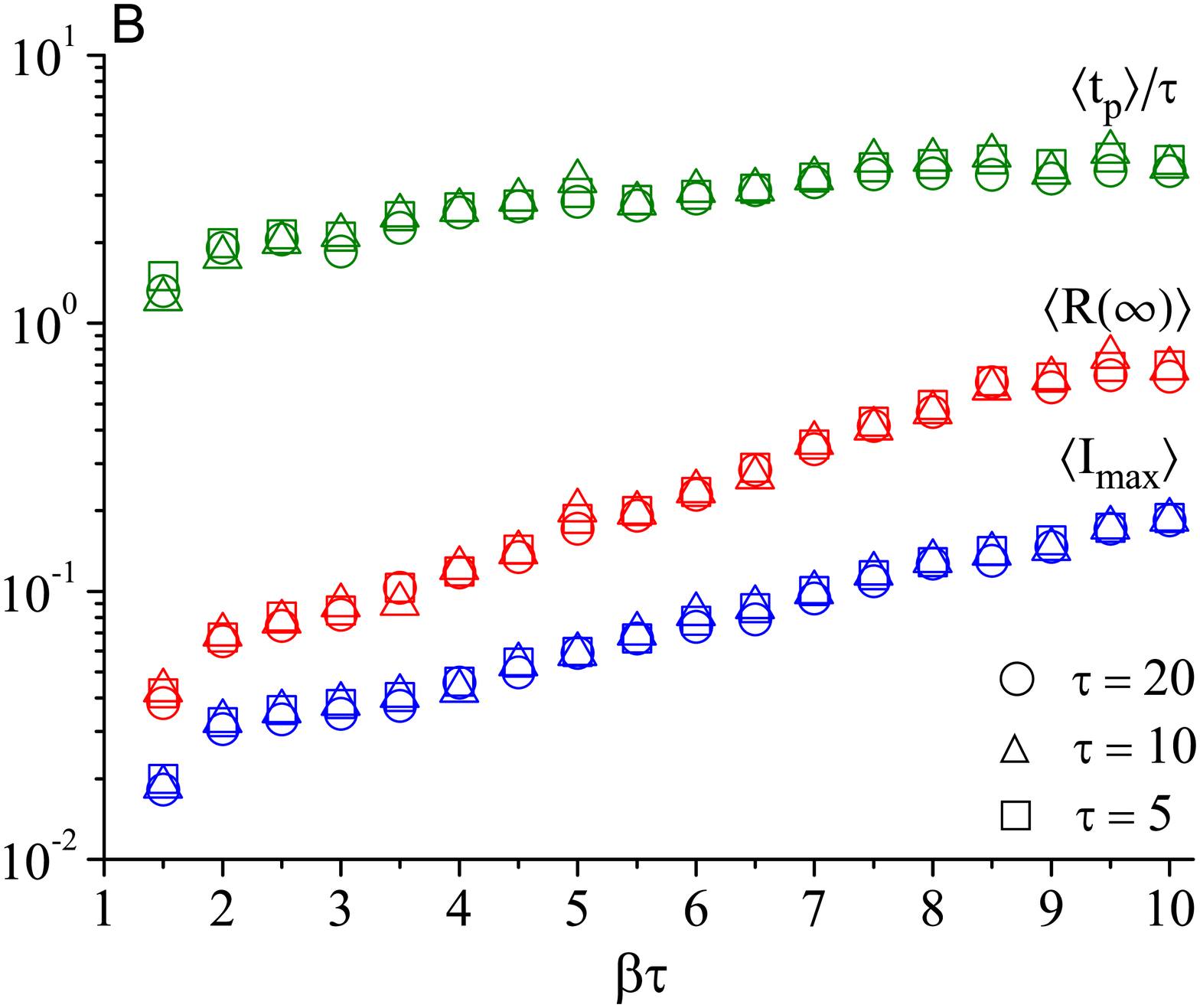}
   }}
\caption{{\bf Behavior of the disease spread in the (static) Romantic
    network regarding:} (A) Comparison between giant component and
  complete network (scaled by $573/288$, i.e. the {\em complete
    network/giant component} size relation) and (B) the $\beta\tau$
  product. Each point of the curves is the result of an average over
  200 independent runs (different seed of the random number
  generator).
}\label{BT}
\end{figure}

Before presenting the results let us make two important comments.  In
the graphical representation of the Romantic network originally
presented in Bearman et al. paper~\cite{Bearman2004}, apart from the
giant component of 288 nodes, there are many small components ranging
from pairs, triplets up to ten nodes which gives a total of 573
students. However any infection started at any one of those small
components will be restricted to it. Consider the whole network of 573
elements results in a theoretical renormalization of the giant
component results by the $288/573$ ratio, which we have checked (see
Fig.~\ref{BT}A). The second comment refers to Figure \ref{BT}B, which
clearly shows that different pairs of $\beta$ and $\tau$ values, but
with the same value of $\beta\tau$, collapse to the same point. This
is well known from the differential equation of the classical SIR
model~\cite{Murray}, however we want to explicitly show that it holds
too for numerical simulations on static networks.  So the statistical
properties of disease outbreaks depend only on the product $\beta\tau$
which, along with the degree distribution $P(k)$, defines the
theoretical value of the basic reproductive number $R_0$.  Therefore,
in what follows, our results will be shown in term of $\beta\tau$
which, as expected, appears to be a good parameter to characterize all
of them.

\begin{figure}[ht!]\centerline{
  \subfigure{\includegraphics[width=6cm]{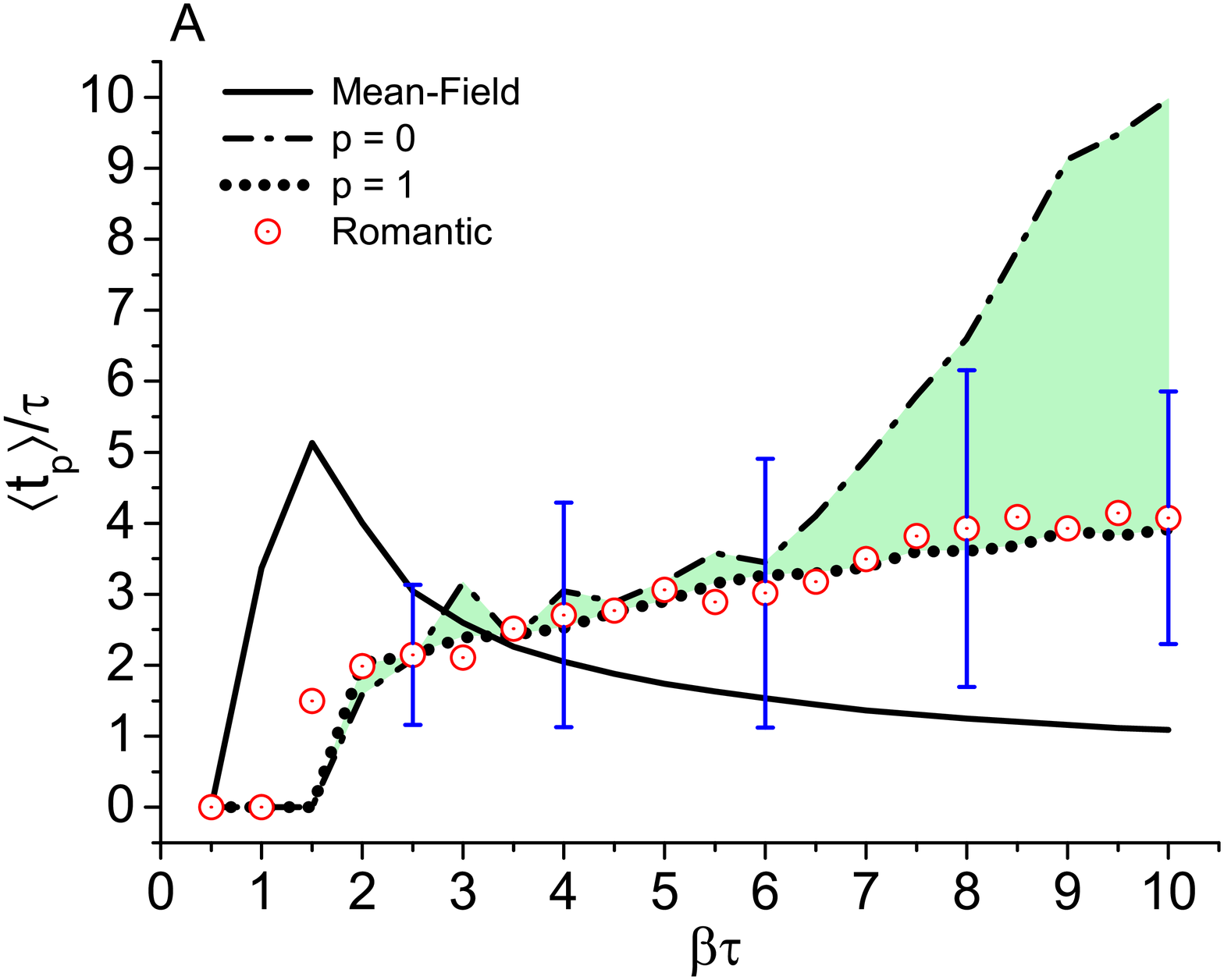}}\hfil
  \subfigure{\includegraphics[width=6cm]{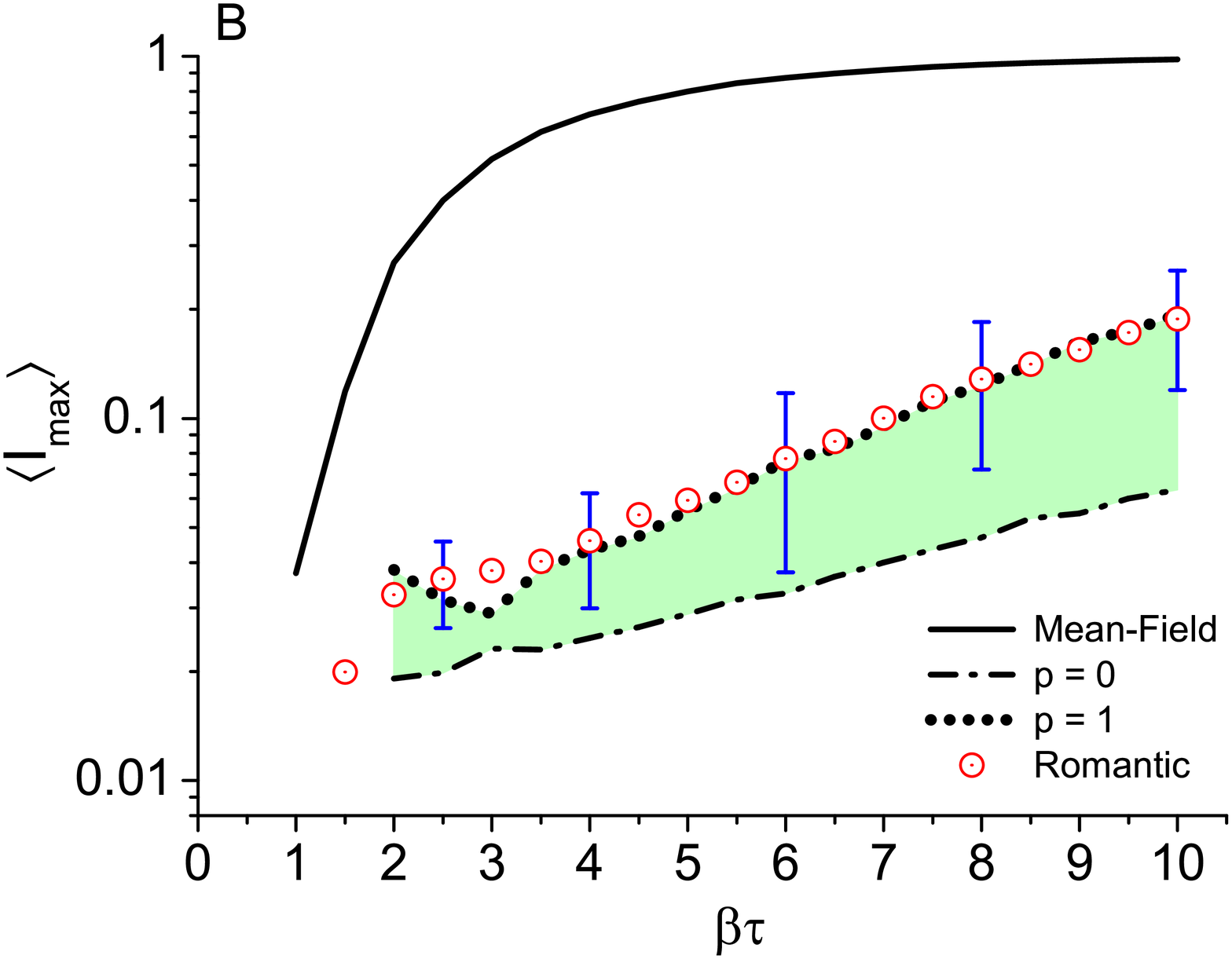}}}
  \centerline{
  \subfigure{\includegraphics[width=6cm]{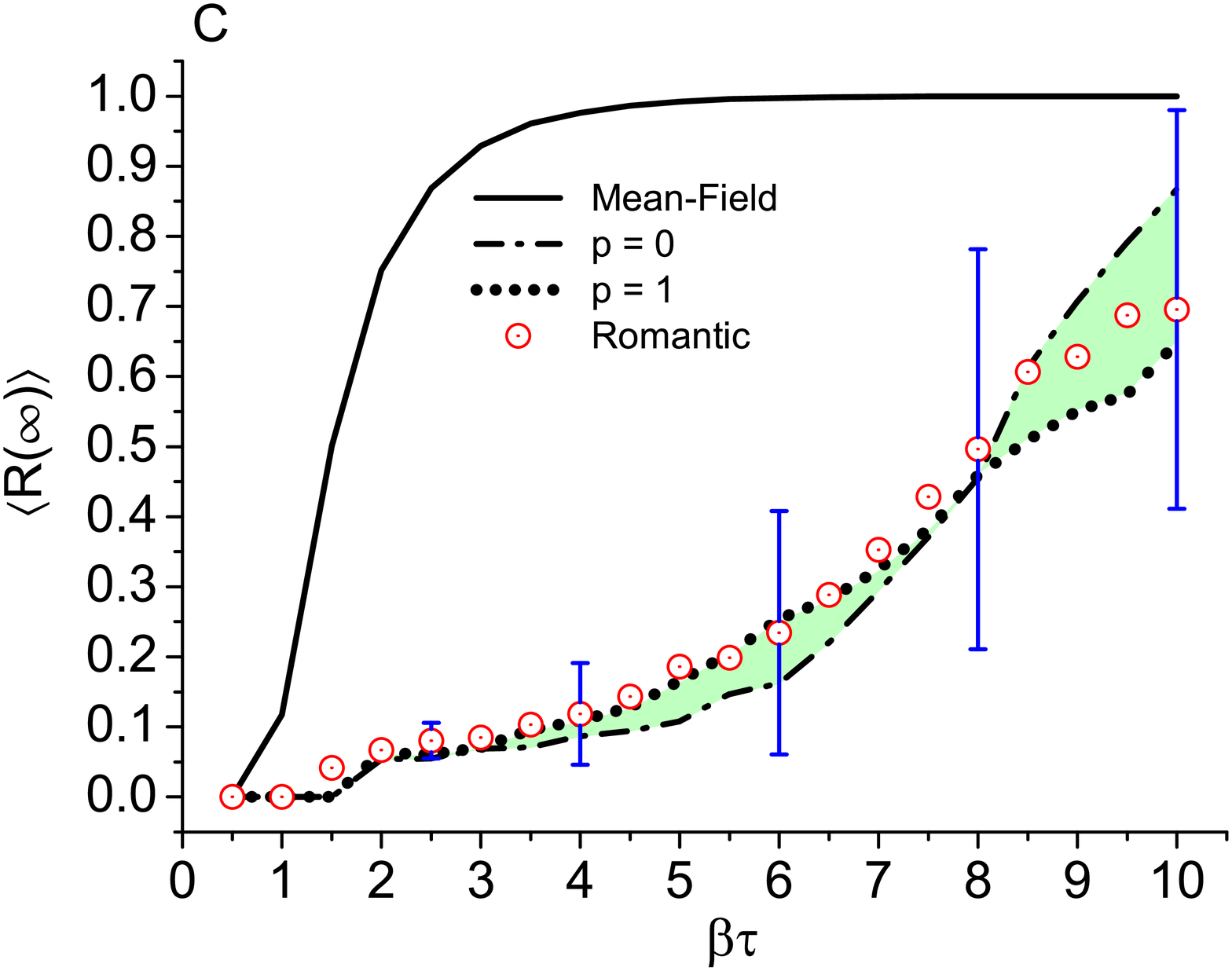}}\hfil
  \subfigure{\includegraphics[width=6cm]{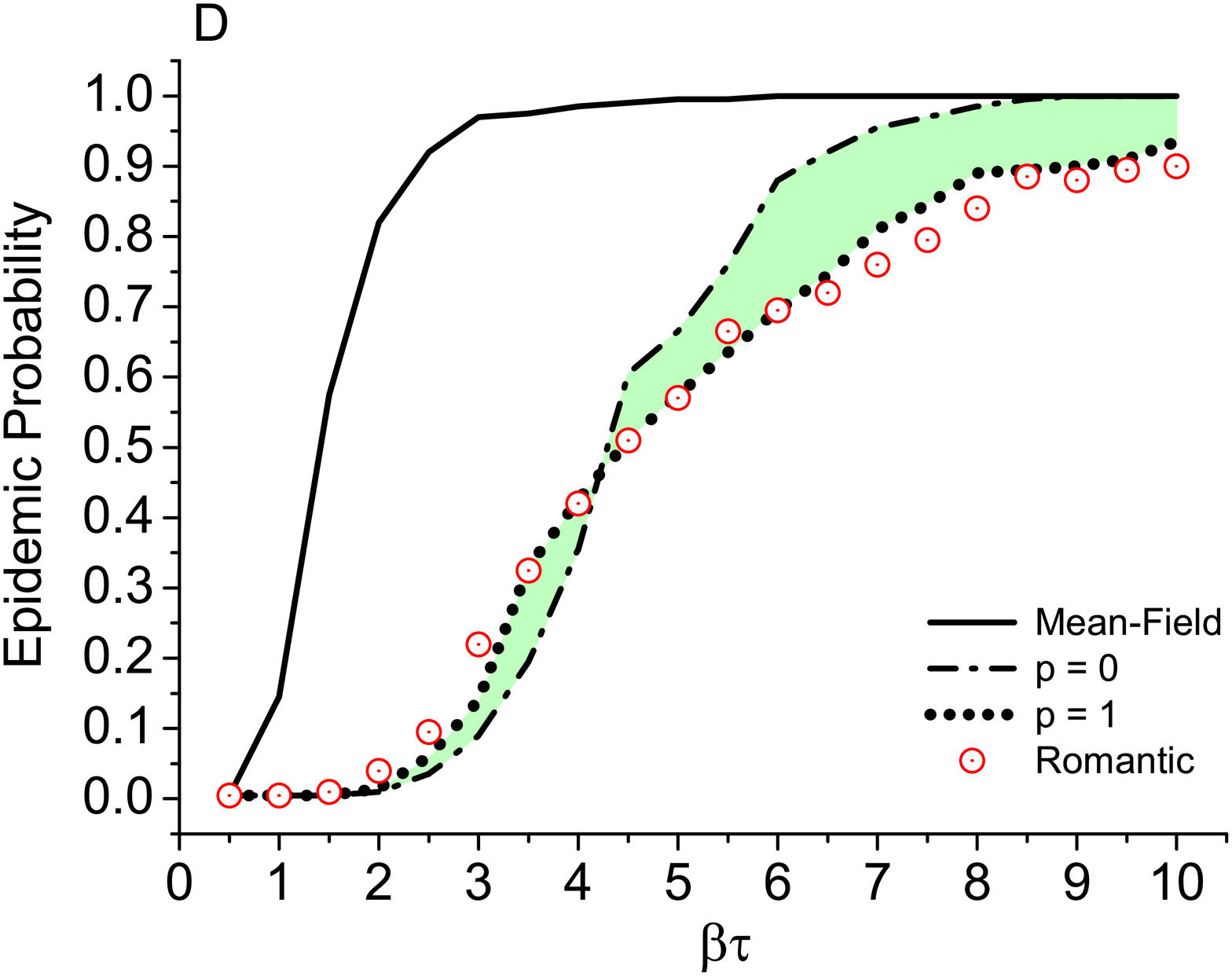}}
  }
  \caption{{\bf Epidemic simulation of the SIR model in different
topologies:} Watts-Strogatz, Romantic, and fully connected
networks. Results for (A) time to the epidemic peak, (B) maximum
number of infectives, (C) total number of removed or final prevalence,
and (D) epidemic probability, defined as the fraction of realizations
that end up with more than 5\% of the population affected. Each point
of the curves is the result of an average over 200 independent runs on
10 different graphs (except for the Romantic network which is only one
graph).  Error bars, displayed only at some points, correspond to the
standard deviation over the 200 runs in the Romantic network case.
Green shadows help to visualize the variation of the outcome due to
the progressive disorder in the W-S networks.
}\label{ws}
\end{figure}

\begin{figure}[ht!]\centerline{
  \subfigure{\includegraphics[width=6cm]{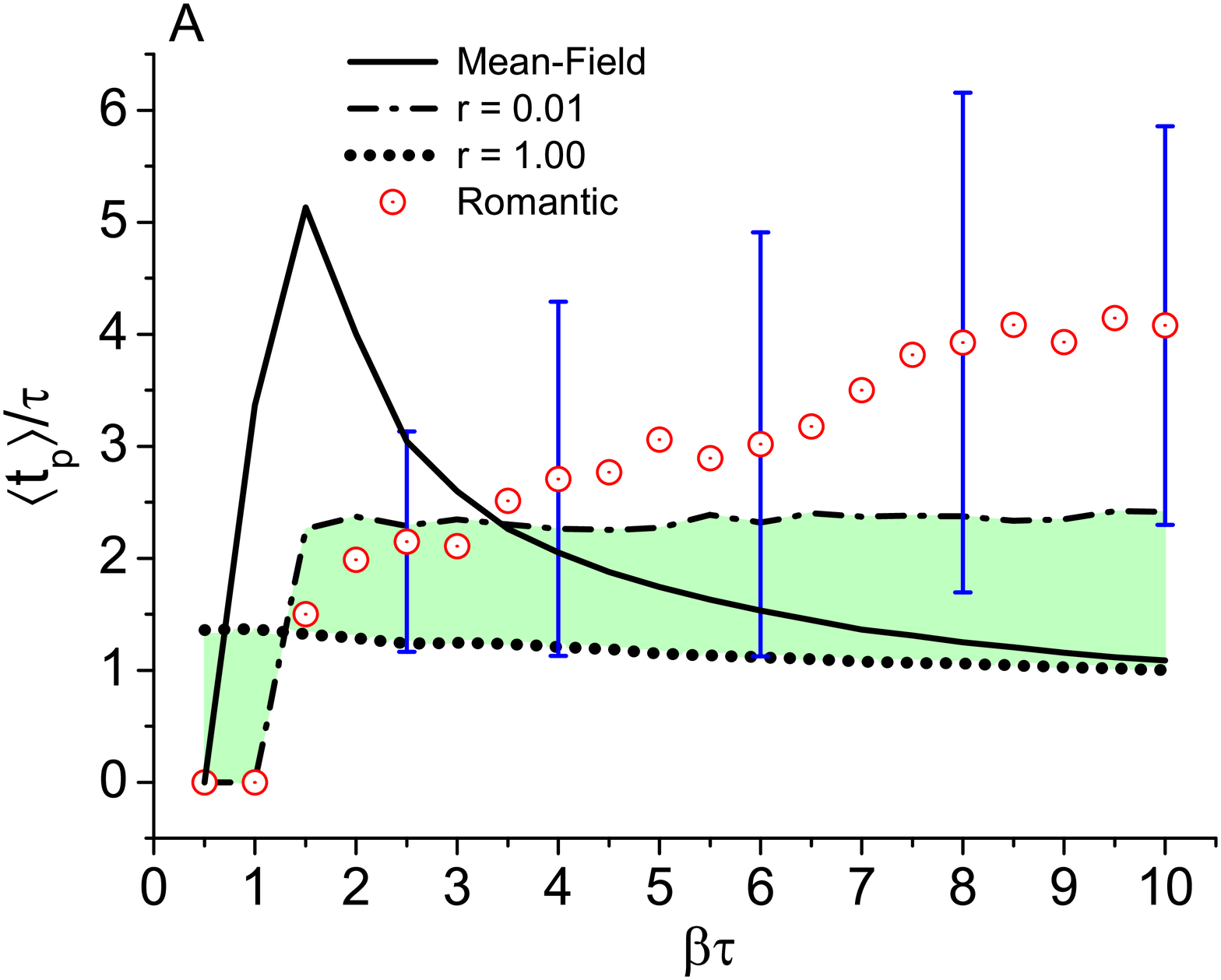}}\hfil
  \subfigure{\includegraphics[width=6cm]{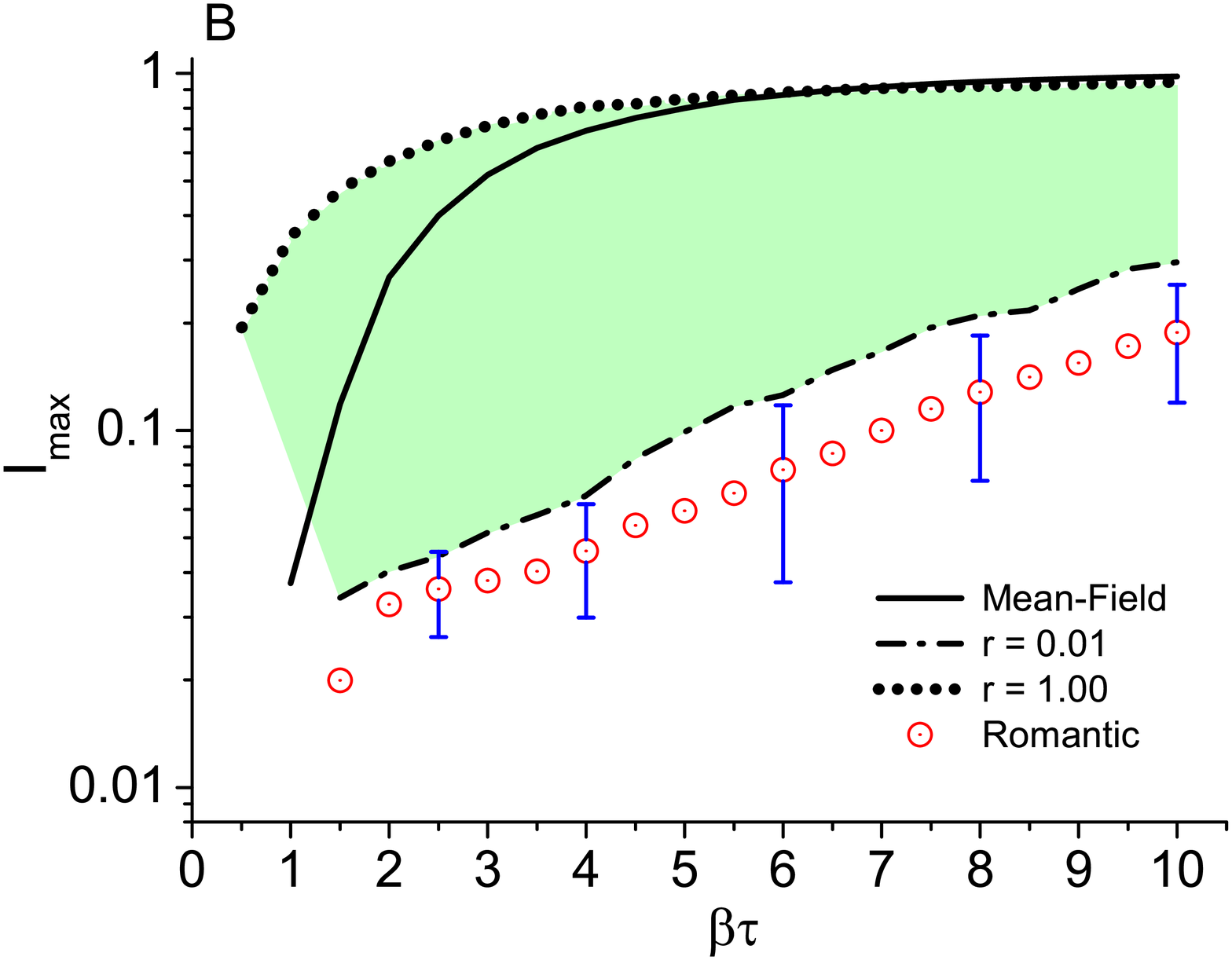}}}
  \centerline{
  \subfigure{\includegraphics[width=6cm]{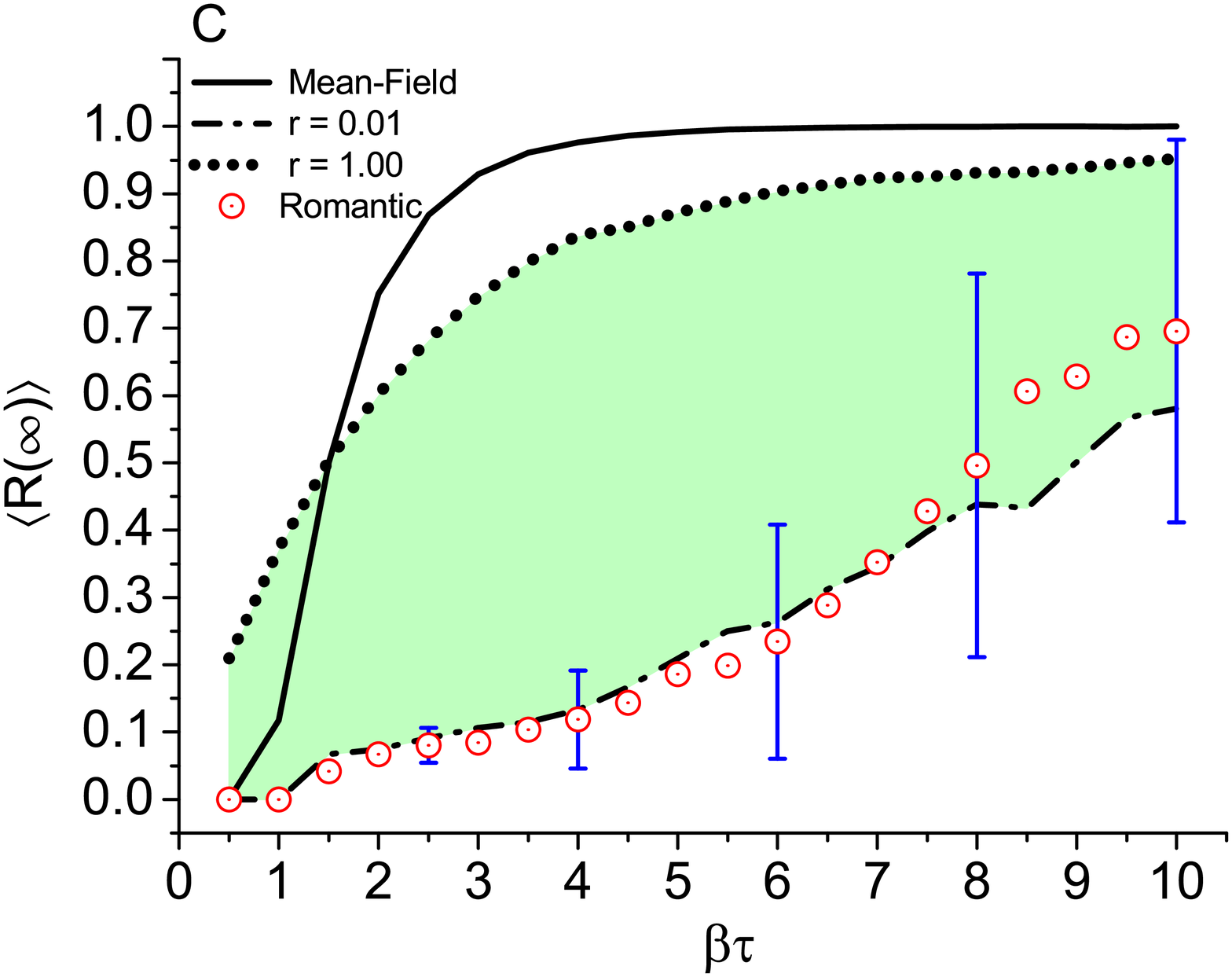}}\hfil
  \subfigure{\includegraphics[width=6cm]{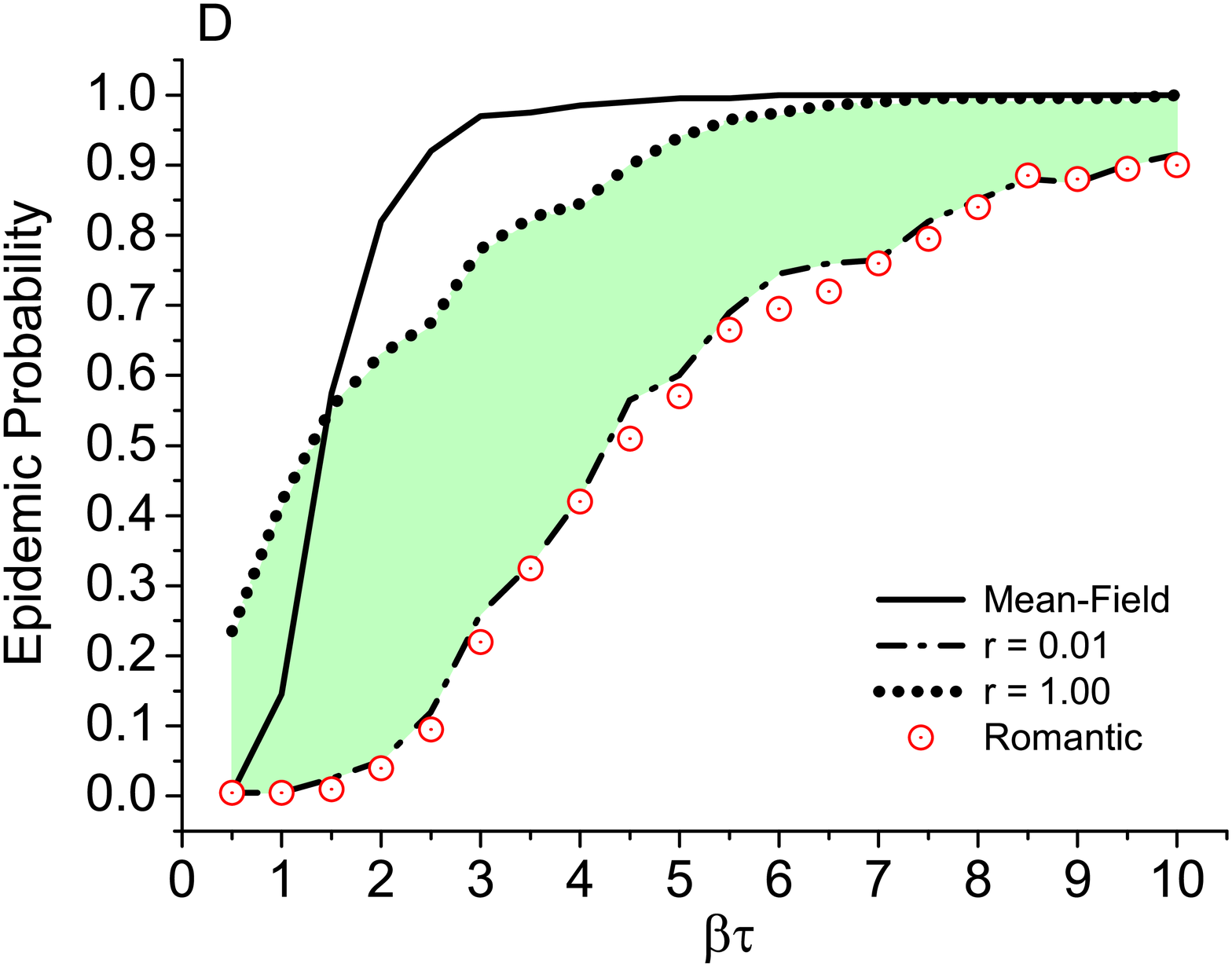}}
  }
  \caption{{\bf Epidemic in different topologies:} Krapivsky-Redner,
Romantic, and fully connected networks. Results for (A) time to the
epidemic peak, (B) maximum number of infectives, (C) final prevalence,
and (D) epidemic probability.  Number of runs, average per point, and
error bars as in Fig.~\ref{ws}.  Green shadows help to visualize the
variation of the outcome due to the progressive heterogeneity in the
K-R networks.
}\label{kr}
\end{figure}

From a general perspective, we observe that, except for the K-R
networks with $r \approx 1$ (the most heterogeneous networks), many
topologies give similar epidemic results provided that $\beta\tau$ is
below $4$. This can be concluded from the comparison between the
Romantic and the W-S networks, as can be seen in Fig.\ref{ws}, and the
K-R for $r < 1$, as shown in Fig.\ref{kr}. Moreover, if we consider a
small group of topologies, the W-S with large $p$ ($\approx 1$) and
the K-R with small $r$ ($\approx 0$), the behavior of the results,
specially $R(\infty)$ and the epidemic probability
(Figs.~\ref{ws},\ref{kr} C,D) are very similar to the corresponding
ones from the Romantic network, up to values of $\beta \tau$ equal to
8.  Considering individual realizations, the displayed error bars in
Fig.\ref{ws}A-C and Fig.\ref{kr}A-C correspond to the Romantic network
and give an idea of the observed dispersion associated to the small
size of it.

Besides, we observe that the mean field (all mixed) scenario is the
worst possible one from the epidemic point of view. Any other
situation, i.e., any other network structure in which not everybody is
potentially connected with anybody, even the K-R heterogeneous
networks yield a less unfavorable picture. The only exception is the
K-R network with $r=1$ (which would give a scale-free network for an
infinite system) but for values of $\beta \tau$ less than $1$.

Comparing the outcome between the different networks and the Romantic
one, it is possible to see that the last one does not fit into any of
the model we have tried. Nevertheless, within the uncertainties of the
fluctuation associated with the small (288 nodes) networks considered
here, an epidemic in a random network obtained with large $p$
($\approx 1$) is statistically undistinguished from an epidemic in the
Romantic network. On the other side, the topology that diverges more
from the Romantic network (and from all the others) is the K-R
network with $r=1$. This particular value or $r$ gives a power law
distribution network with exponent $\gamma=2$ (scale-free) in the
limit of large sizes.  In the present case, however, due to the small
size of the networks, results in a star topology, very different from
the ring-like type of the Romantic network.

\subsubsection*{External field}
The Romantic network was constructed taking into account all the
sexual contacts of the surveyed school but excluding any possible
contact outside the school. This section is devoted to explore the
effect that a possible contact of a member of the network with an
infected subject outside of it can have in the epidemic dynamics
inside the school.  This effect is controlled by the parameter $B$
which represent the probability of the external excursion to happen.

\begin{figure}[ht!]\centerline{
  \subfigure{\includegraphics[width=5.55cm]{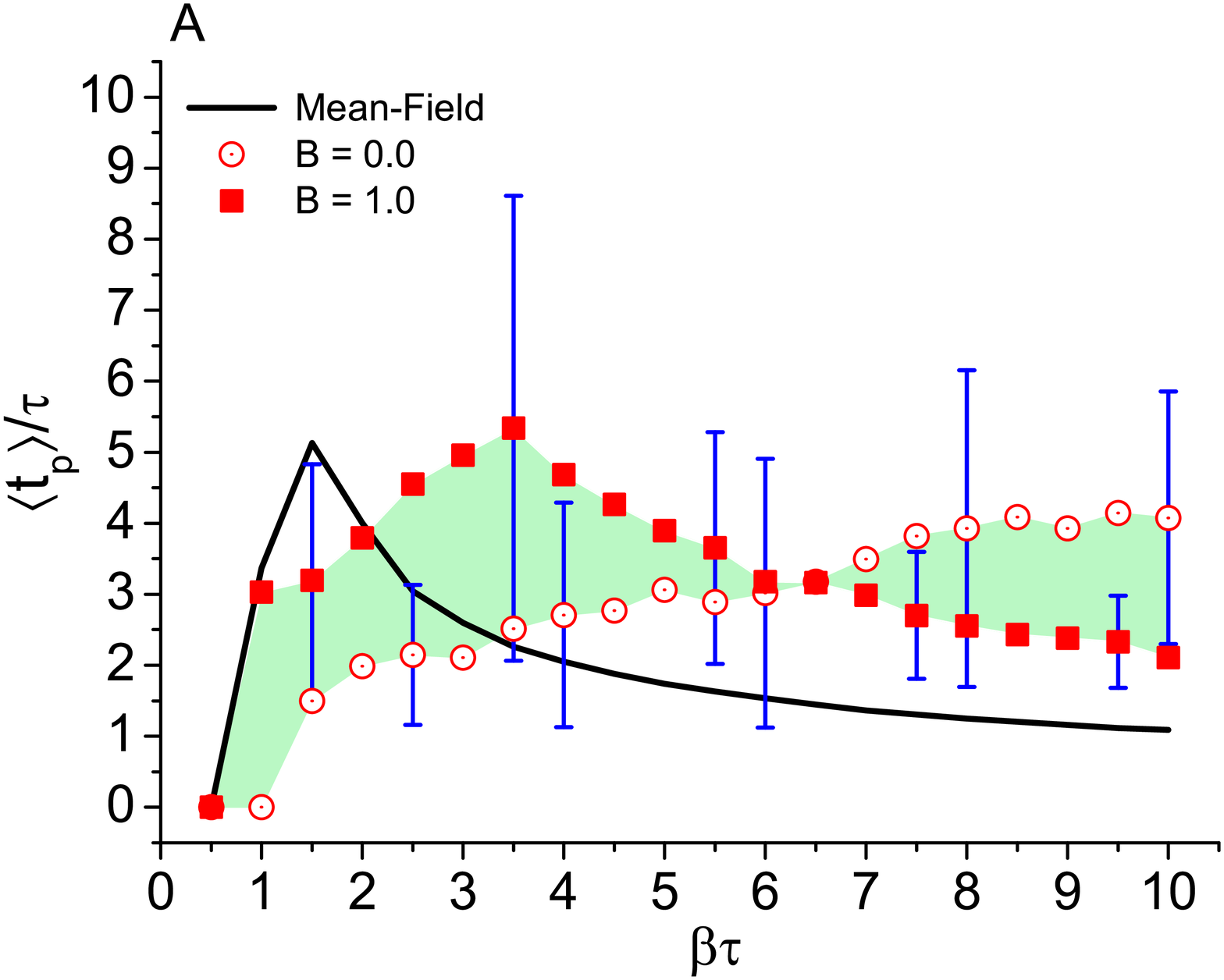}}\hfil
  \subfigure{\includegraphics[width=5.55cm]{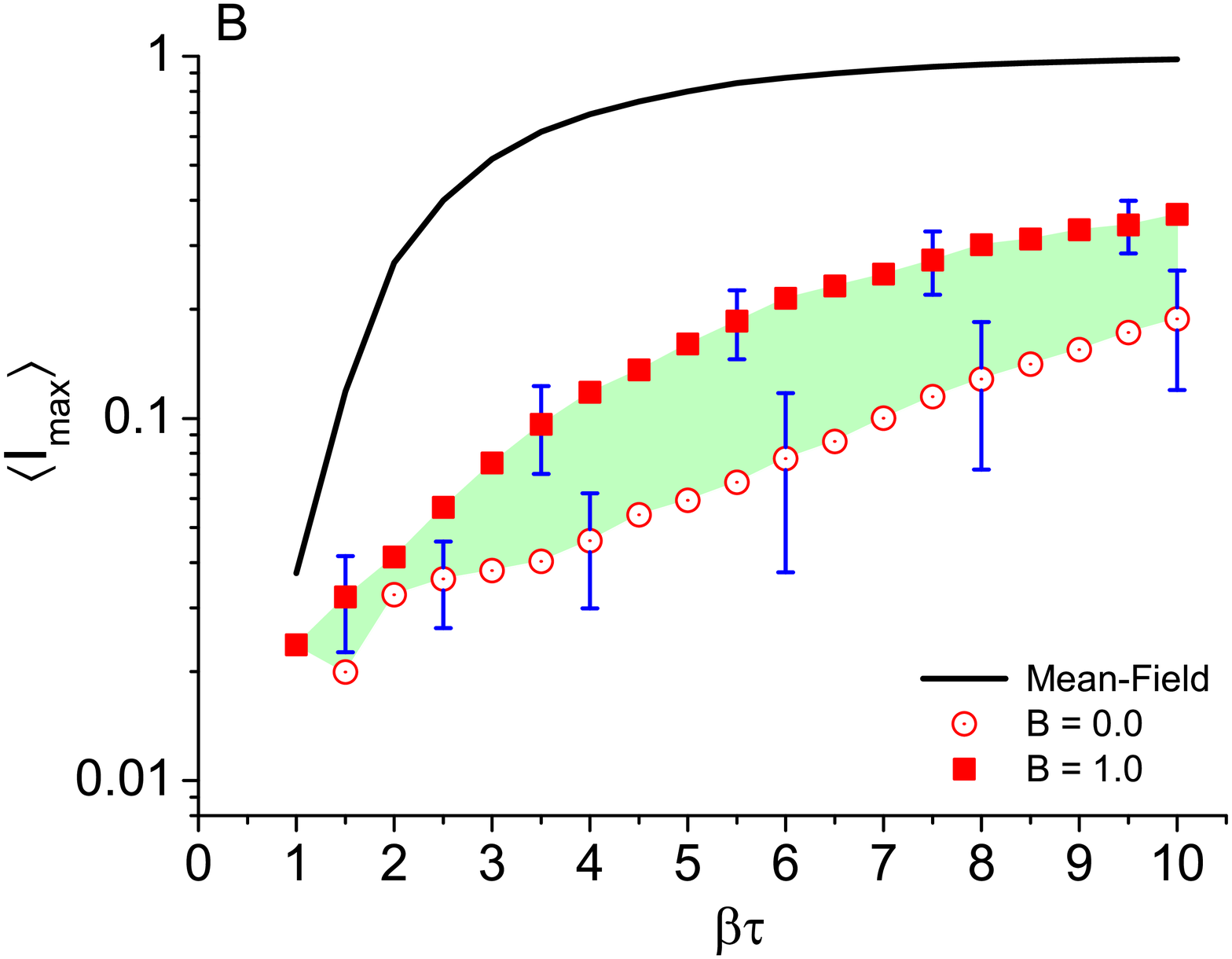}}}
  \centerline{
  \subfigure{\includegraphics[width=5.55cm]{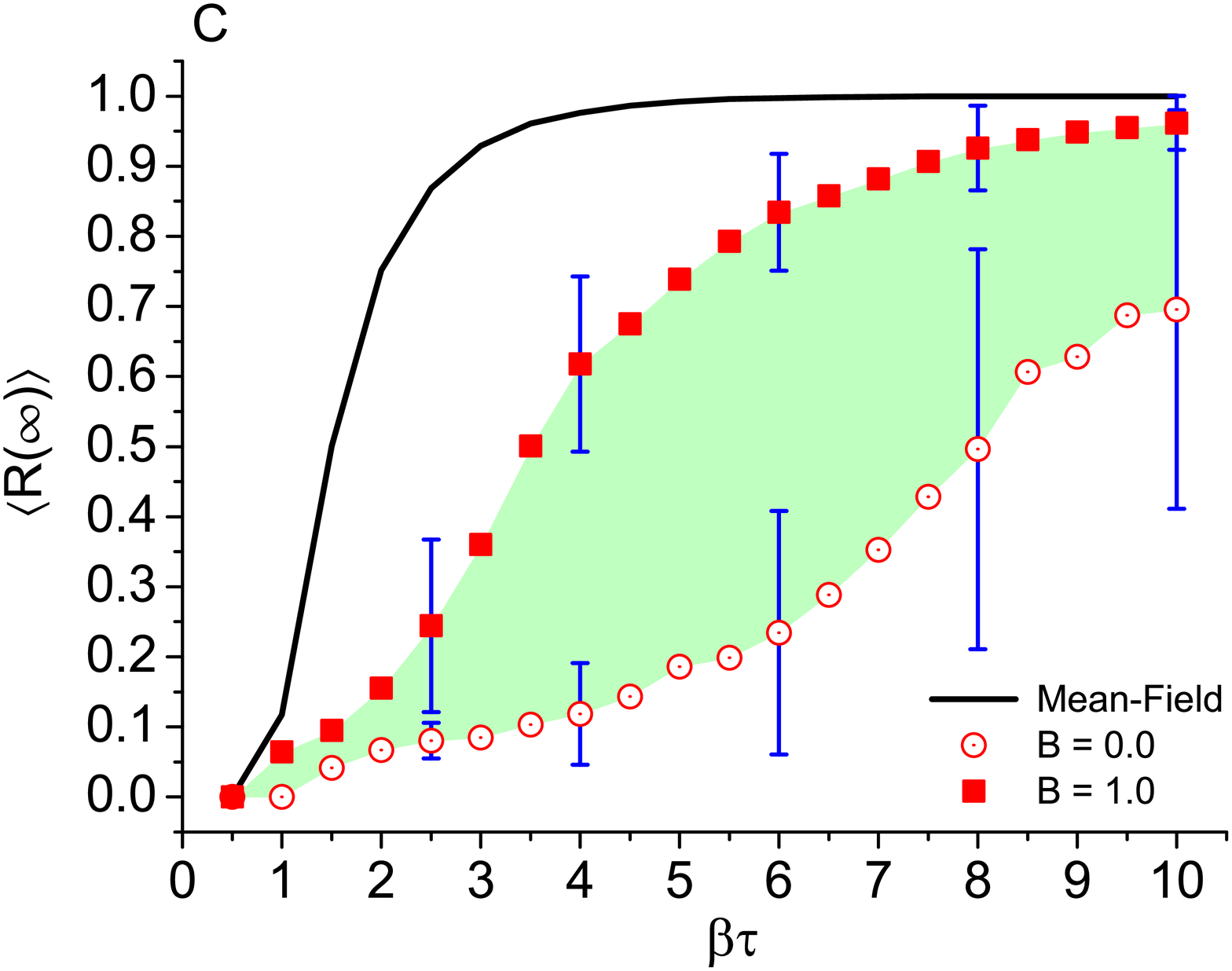}}\hfil
  \subfigure{\includegraphics[width=5.55cm]{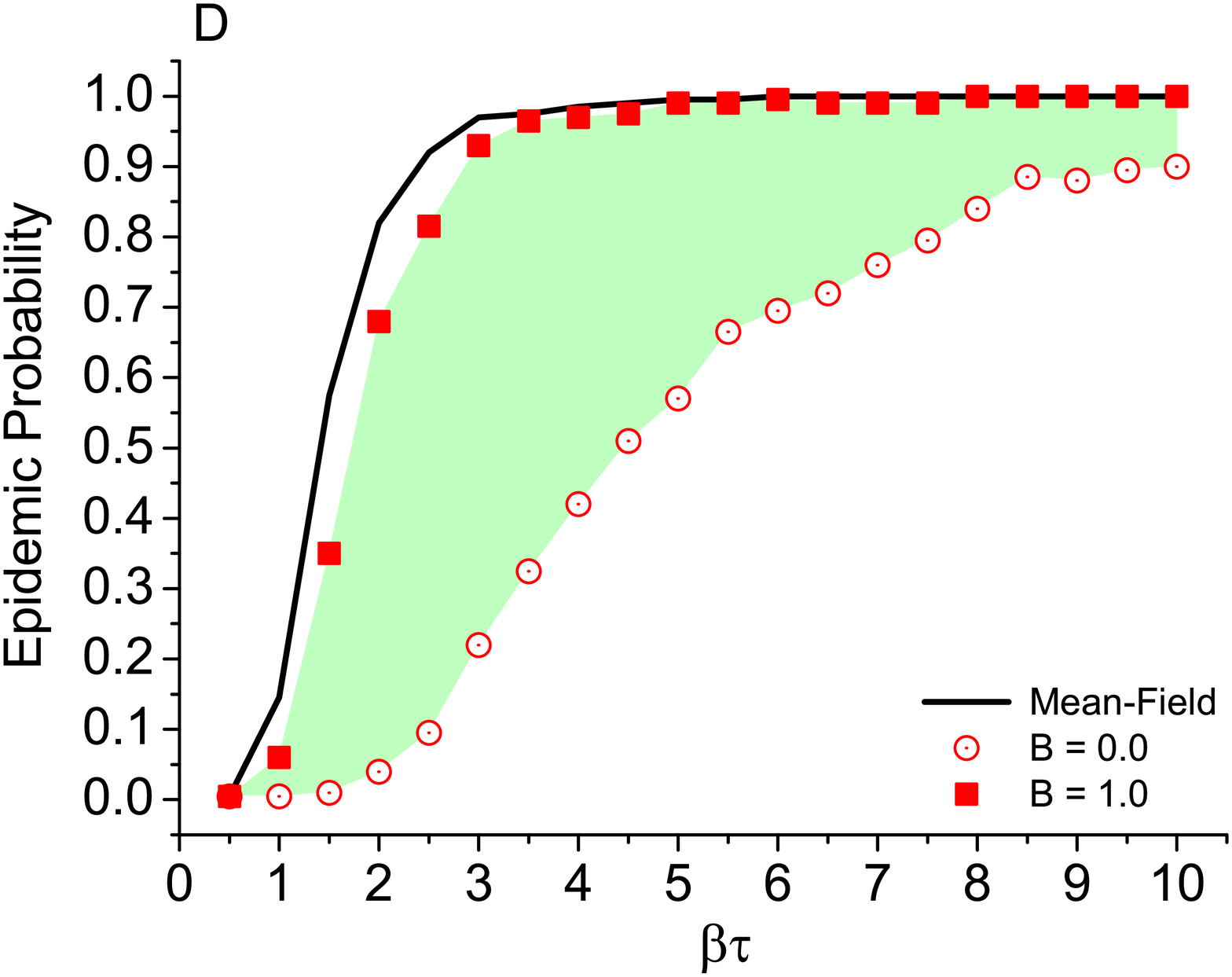}}
  }
  \caption{{\bf Effect of the ``external field'', $B$, applied to the
      Romantic network:} Results for (A) time to the epidemic peak,
    (B) maximum number of infectives, (C) final prevalence, and (D)
    epidemic probability. Comparison with standard mean field results
    in a network of the same size.  Number of runs, average per point,
    and error bars as in Fig.~\ref{ws}.  Green shadows help to
    visualize the variation of the outcome as increasing the $B$
    parameter.}
\label{campo}
\end{figure}

As can be see in Fig.~\ref{campo}, the effect of the external field
makes the outcome of the infection, measured by the total number of
infected individuals as a function of $\beta \tau$, closer to what is
expected from the standard SIR model.  This suggests that the
practical effect of the external field is to make the network closer
to a full connected one.

\subsubsection*{Degree of the initial infected}
\begin{figure}[ht!]\centerline{
 \subfigure{\includegraphics[width=5.55cm]{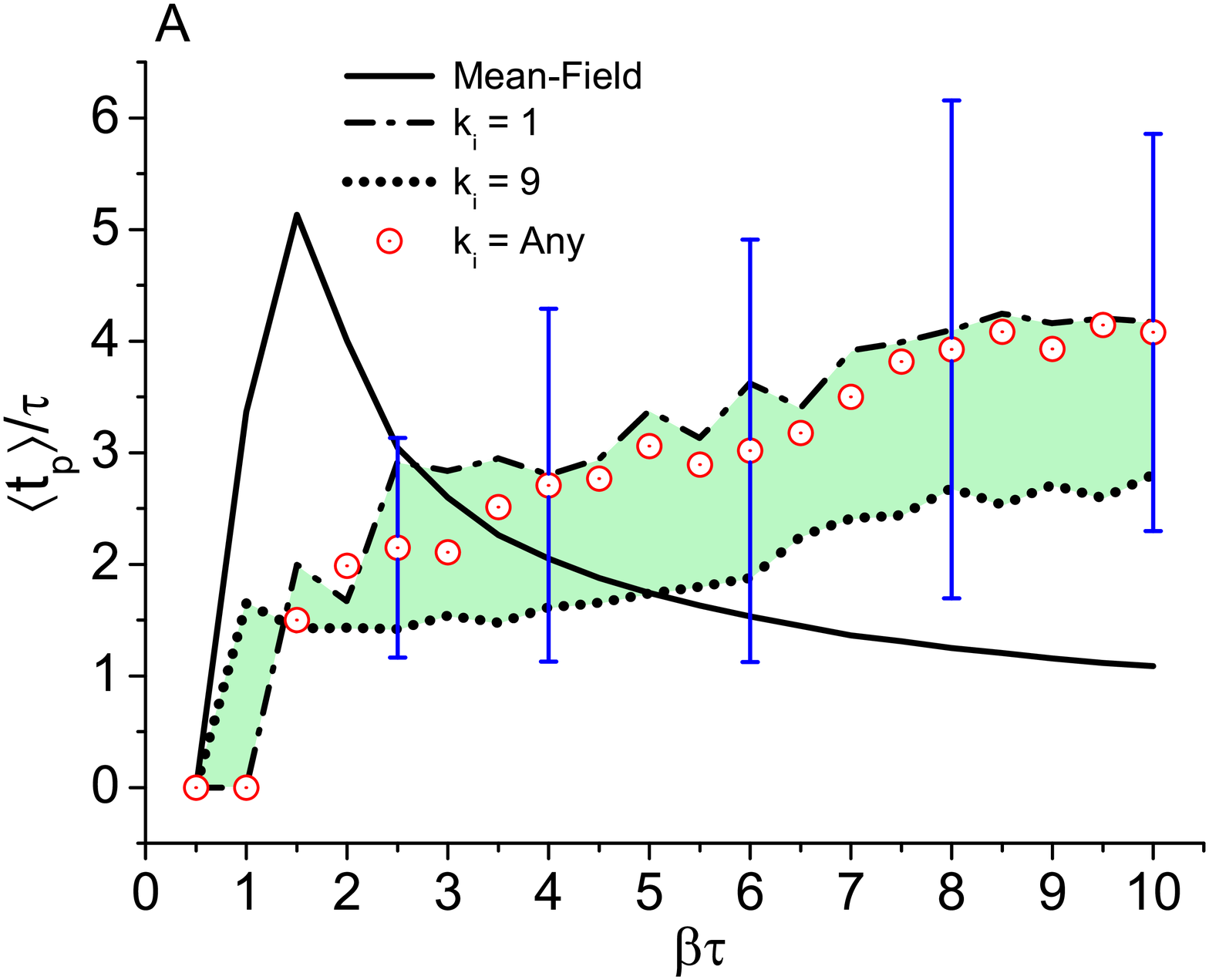}}\hfil
  \subfigure{\includegraphics[width=5.55cm]{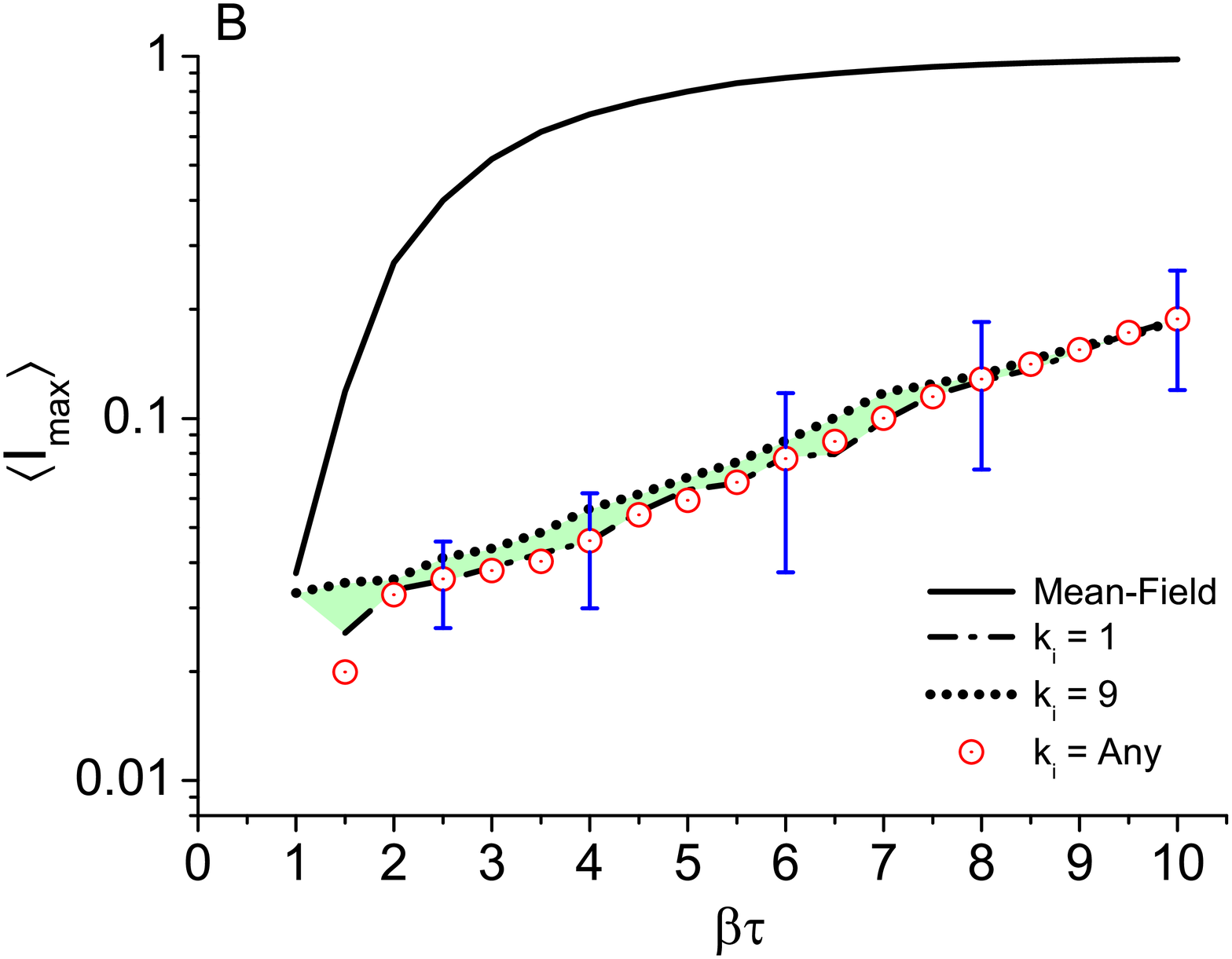}}}
  \centerline{
  \subfigure{\includegraphics[width=5.55cm]{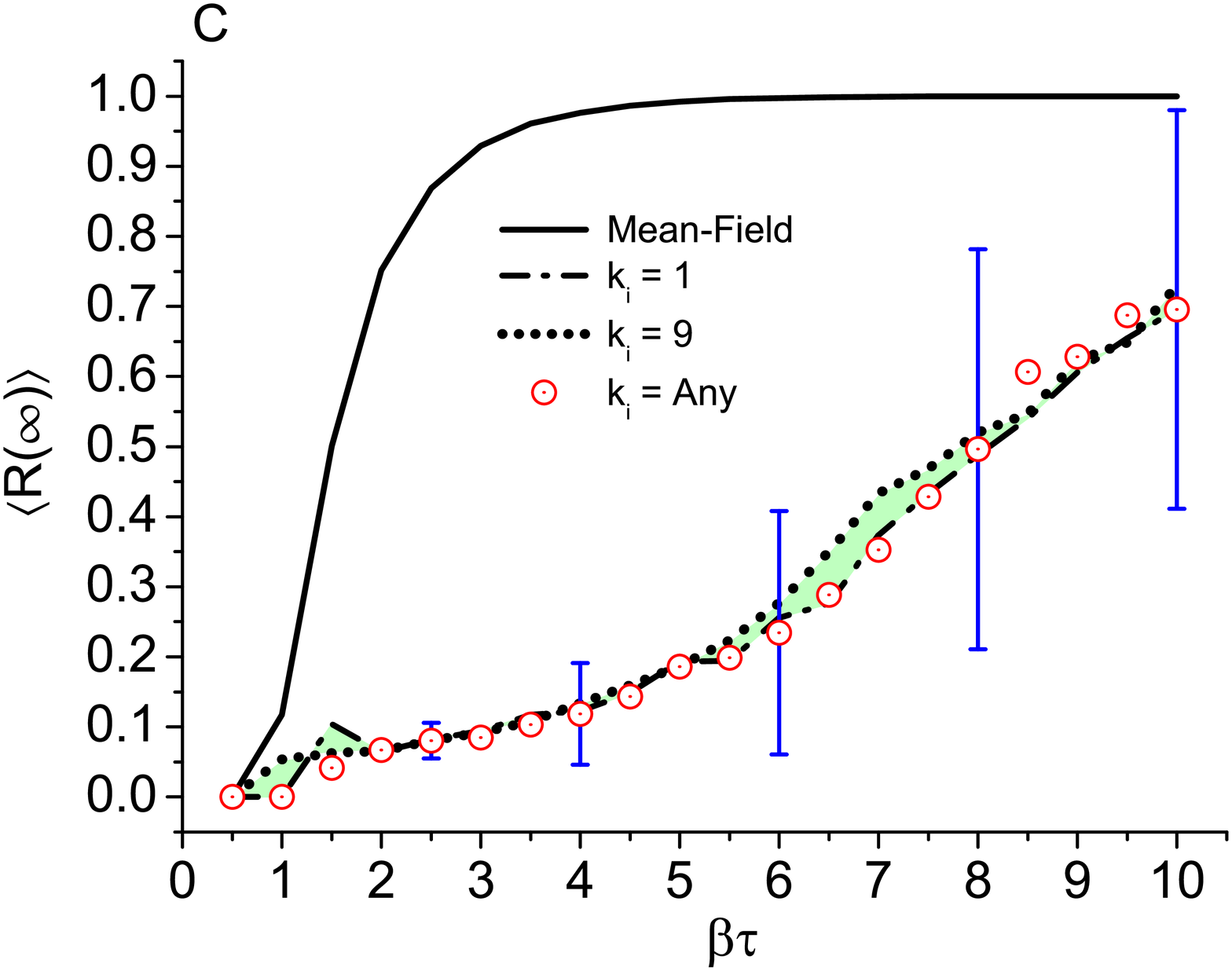}}\hfil
  \subfigure{\includegraphics[width=5.55cm]{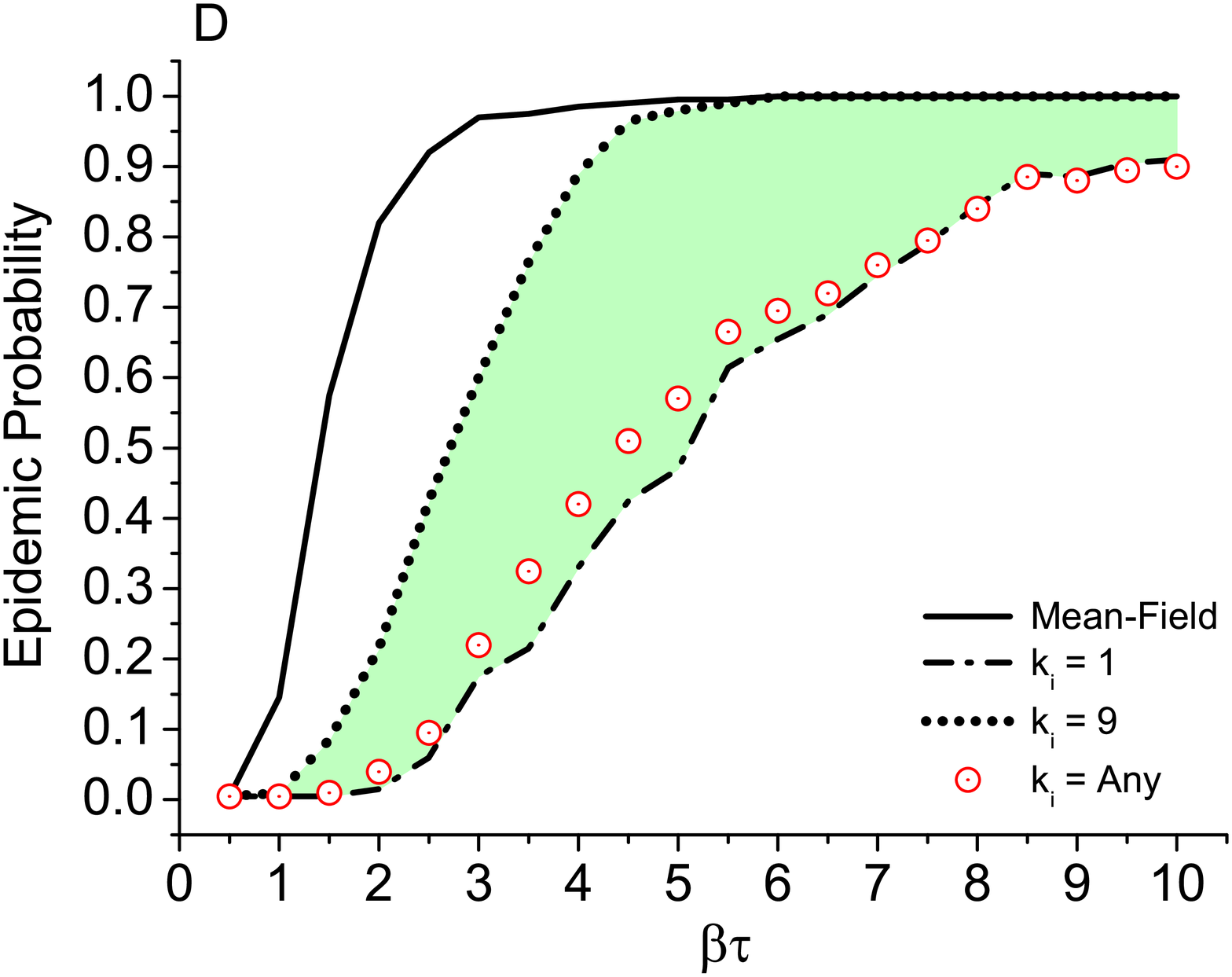}}
  }
  \caption{{\bf Effect of degree of the subject chosen to be the
      initial infected in the Romantic network:} Results for (A) time
    to the epidemic peak, (B) maximum number of infectives, (C) final
    prevalence, and (D) epidemic probability.  Comparison with
    standard mean field results in a network of the same size.  Number
    of runs, average per point, and error bars as in Fig.~\ref{ws}.
    Green shadows help to visualize the variation of the outcome as
    response to the degree variation of the initial infected subject.
}\label{promis}
\end{figure}

The results, presented in Fig.~\ref{promis}, are a little subtle to
interpret.  Looking at Fig.~\ref{promis}B,C ($I_{max}$ and $R(\infty)$
respectively) we are tempted to conclude that the influence of whom is
the initial infected one, if it is more o less connected, is
irrelevant ---which we have to admit is not obvious. However, the
other two figures (Fig.~\ref{promis}A,D), representing the time to the
infection peak and specially the last one, i.e., the epidemic
probability, show a clear divergence from low ($k_i=1$) to high
($k_i=9$) degree of the initial infected.  At $\beta\tau=4$ for
example, we clearly see in Fig.~\ref{promis}C that, independently of
who was the initial infected, the average removed subjects at the end
is around $10\%$. Nevertheless, Fig.~\ref{promis}D remind us that in
the low degree case that situation happens only in 30\% of the
instances, contrasting with the 90\% observed in the high degree
case. Therefore, as expected, the degree of the initial infected is a
key factor in the aftermath of a disease propagation in the Romantic
network.

\subsubsection*{Rules of interaction}
\begin{figure}[ht!]
\centerline{
 \subfigure{\includegraphics[width=6cm]{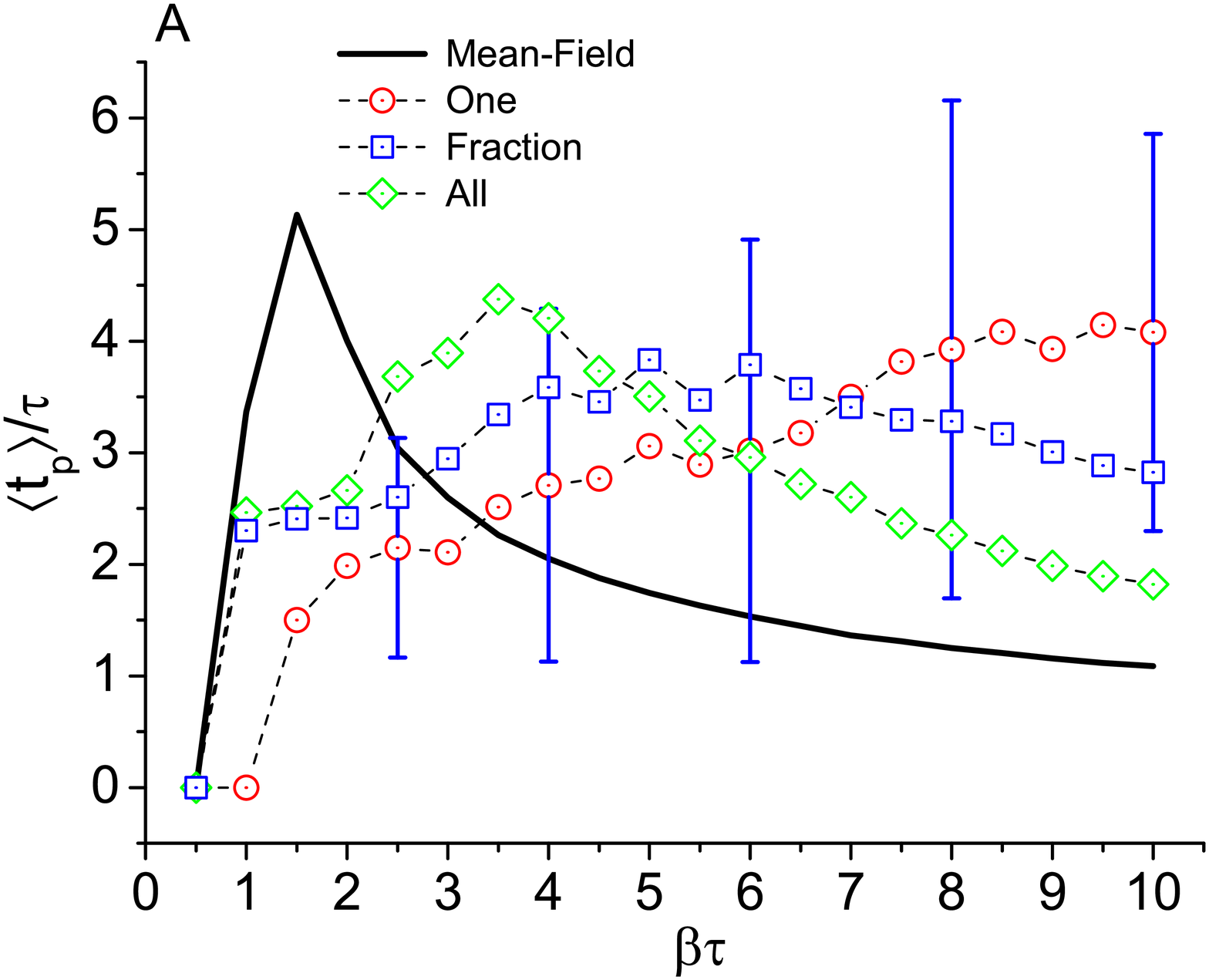}}\hfil
  \subfigure{\includegraphics[width=6cm]{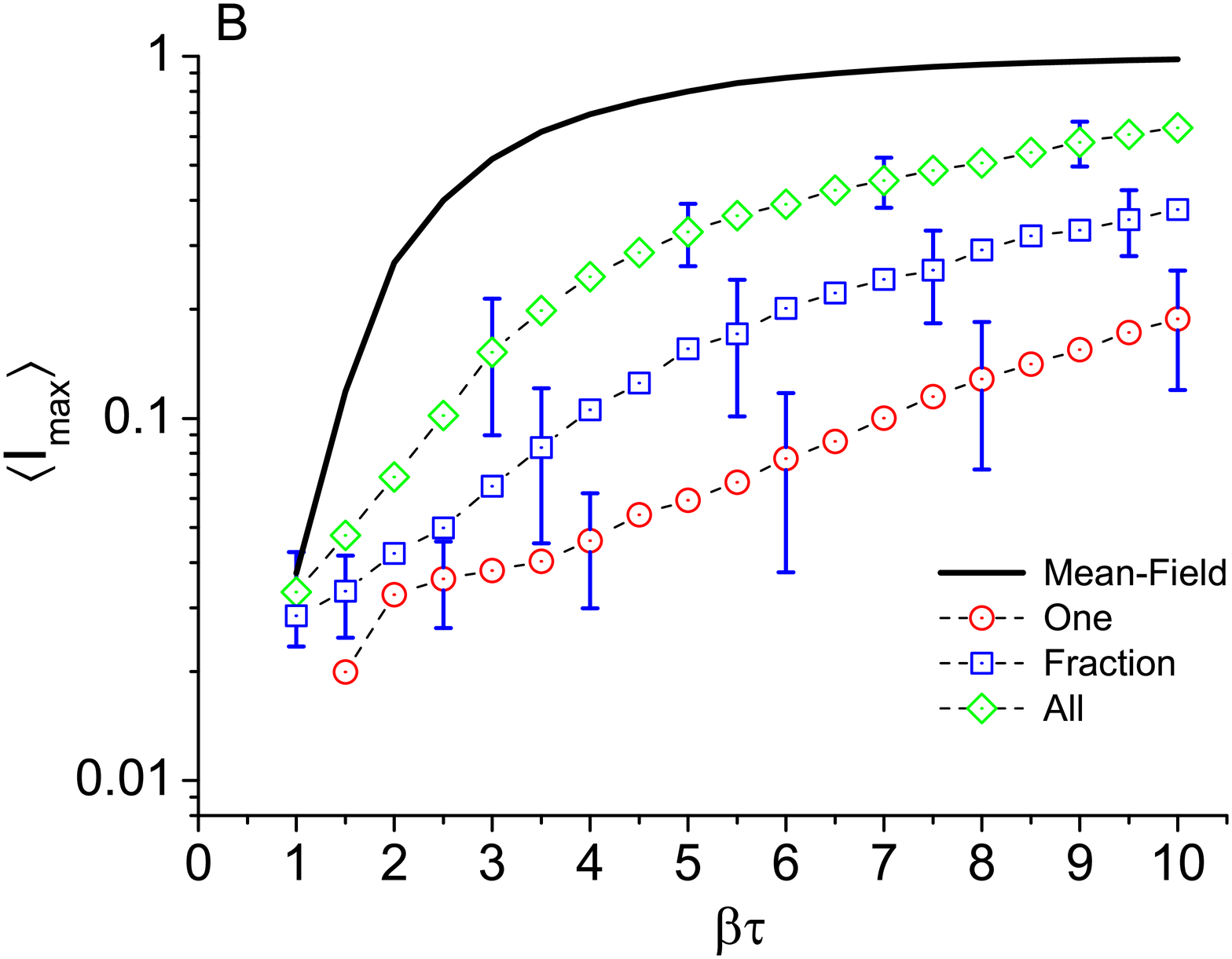}}}
  \centerline{
  \subfigure{\includegraphics[width=6cm]{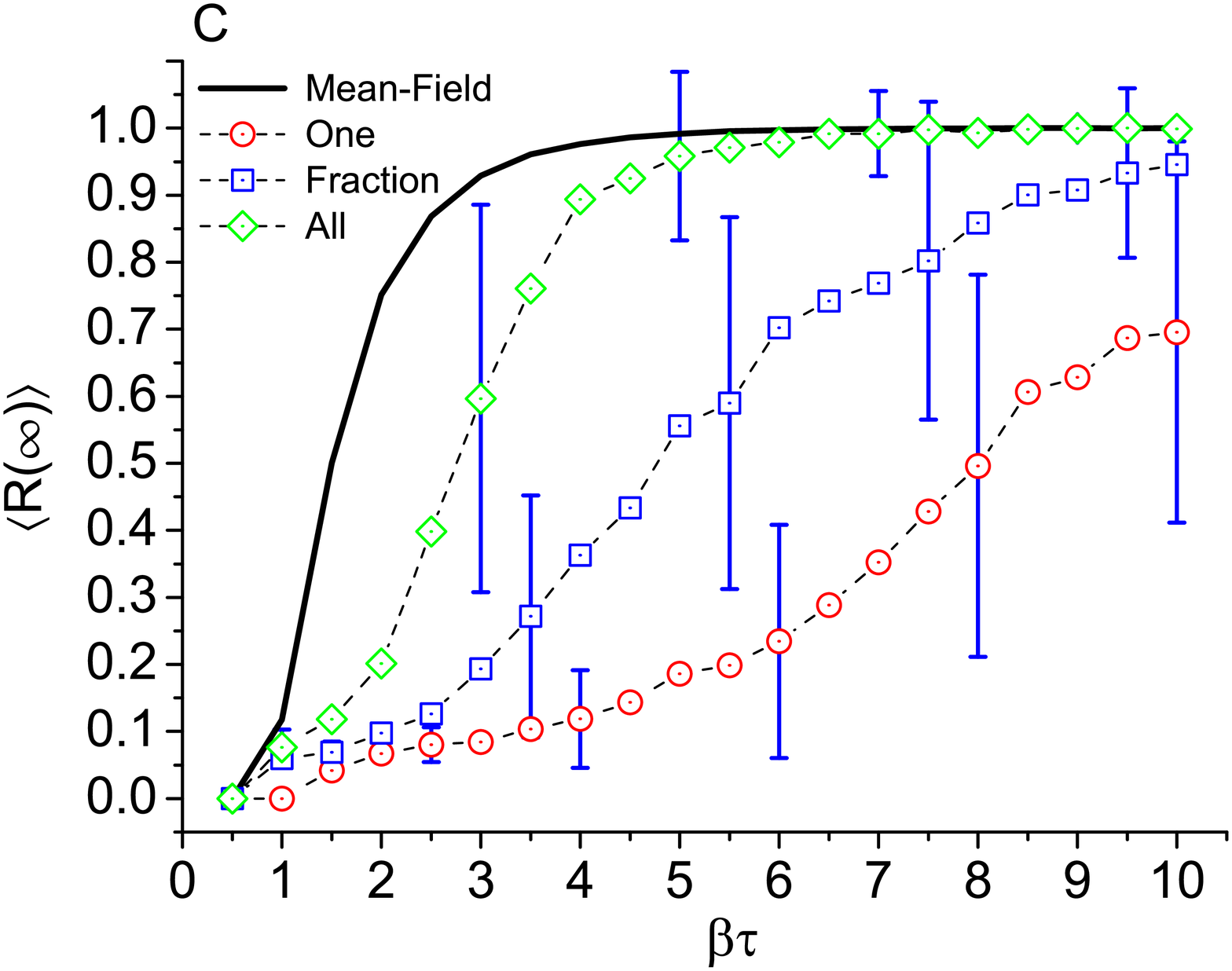}}\hfil
  \subfigure{\includegraphics[width=6cm]{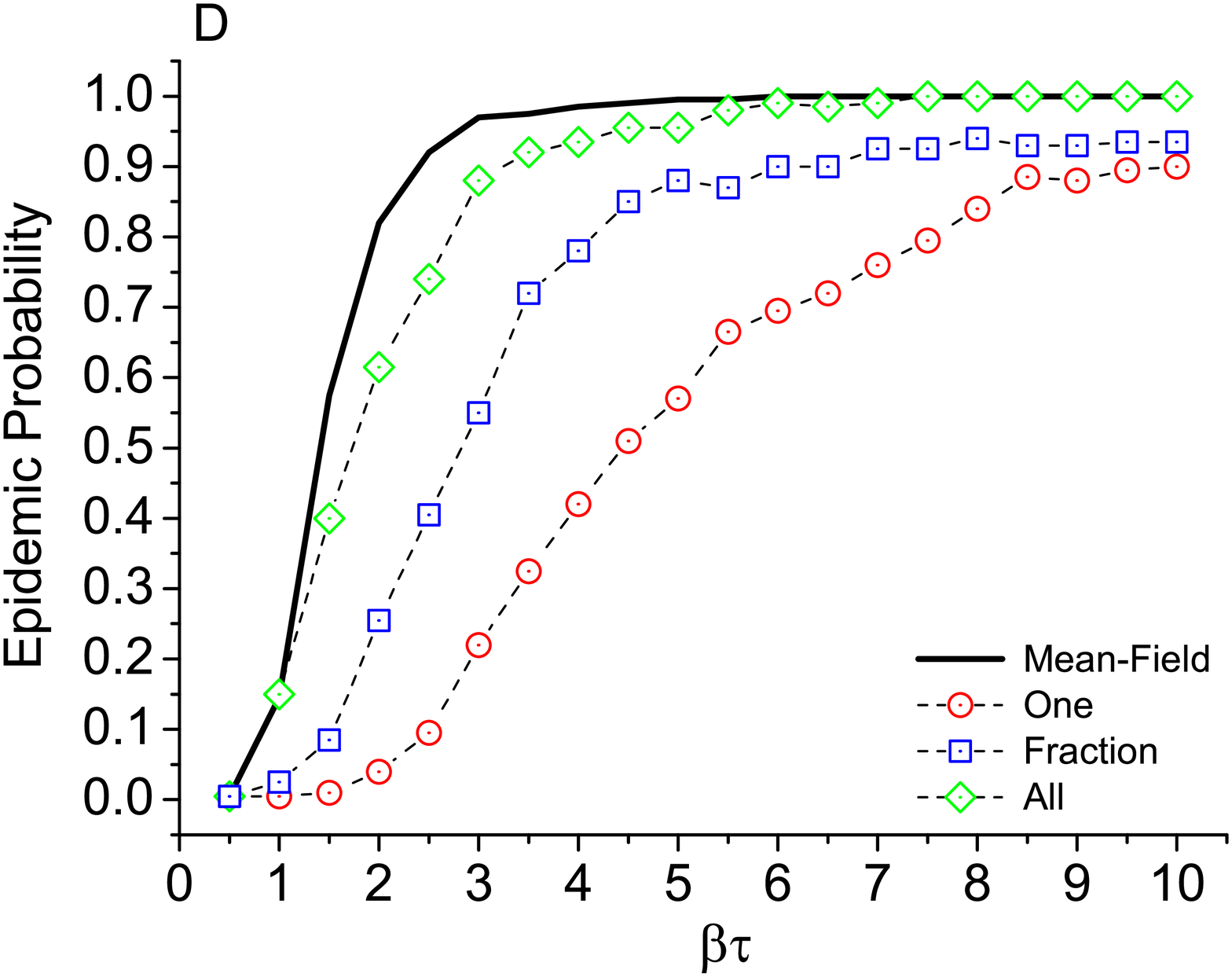}}
  }
  \caption{{\bf Effect of the different rules of interaction on:}
    Results for (A) time to the epidemic peak, (B) maximum number of
    infectives, (C) final prevalence, and (D) epidemic probability.
    Comparison with standard mean field results in the same
    network. Number of runs, average per point, and error bars as in
    Fig.~\ref{ws}.}
\label{fig:rule}
\end{figure}

The three different measures shown in Fig.~\ref{fig:rule} all point
out to the same general trend: from rule ``one'' to rule ``all'' we
gradually approach the mean field all mixed case, which remains as the
upper limit. Even the extreme ``all'' rule is below that limit,
because subjects in the romantic network are not fully connected.

\subsection*{Dynamic Scenario}

\subsubsection*{Measurements}
\begin{figure}[htb!]
\centerline{
\subfigure{\includegraphics[width=6cm]{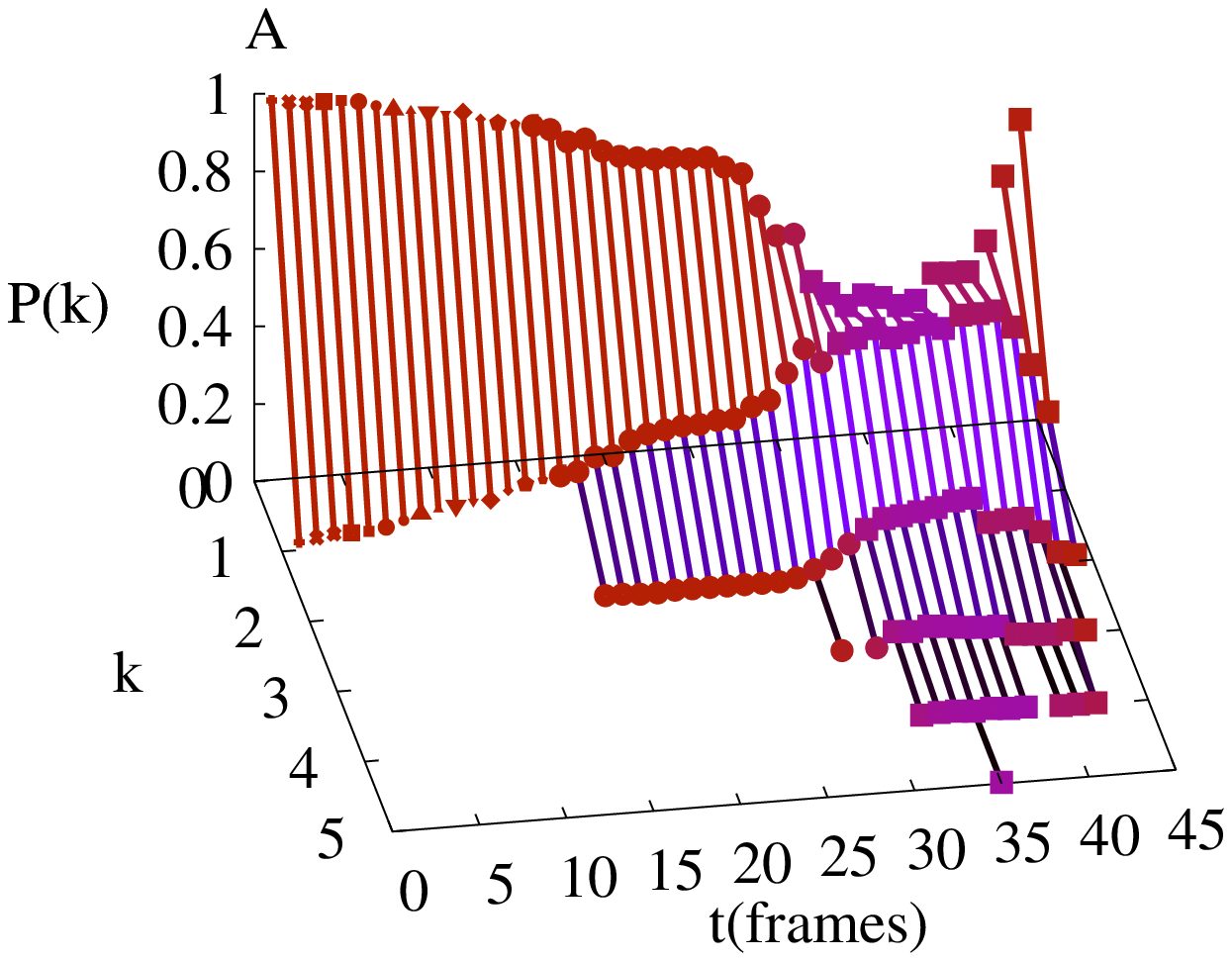}
   } \hfil
\subfigure{\includegraphics[width=6cm]{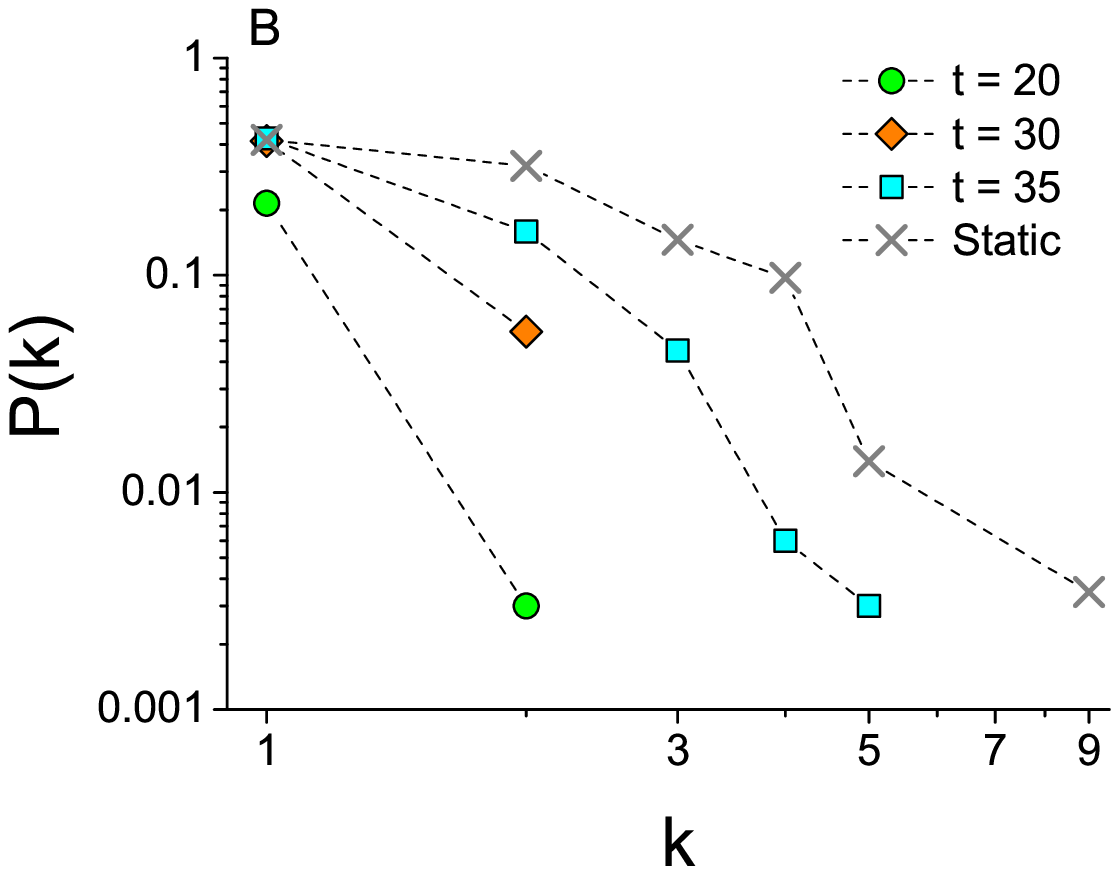}
  } }
   \centerline{
\subfigure{\includegraphics[width=6cm]{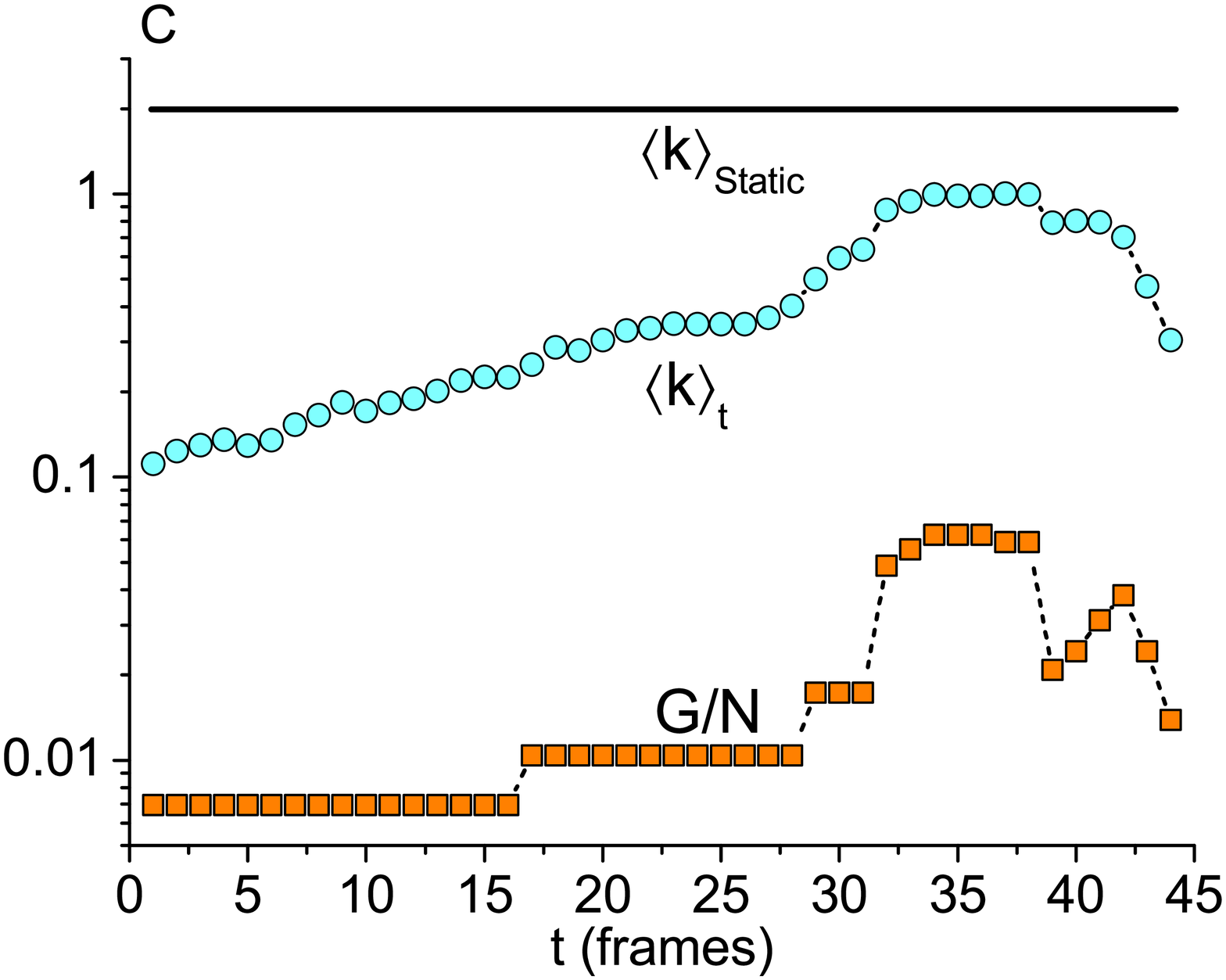}
   } \hfil
\subfigure{\includegraphics[width=6cm]{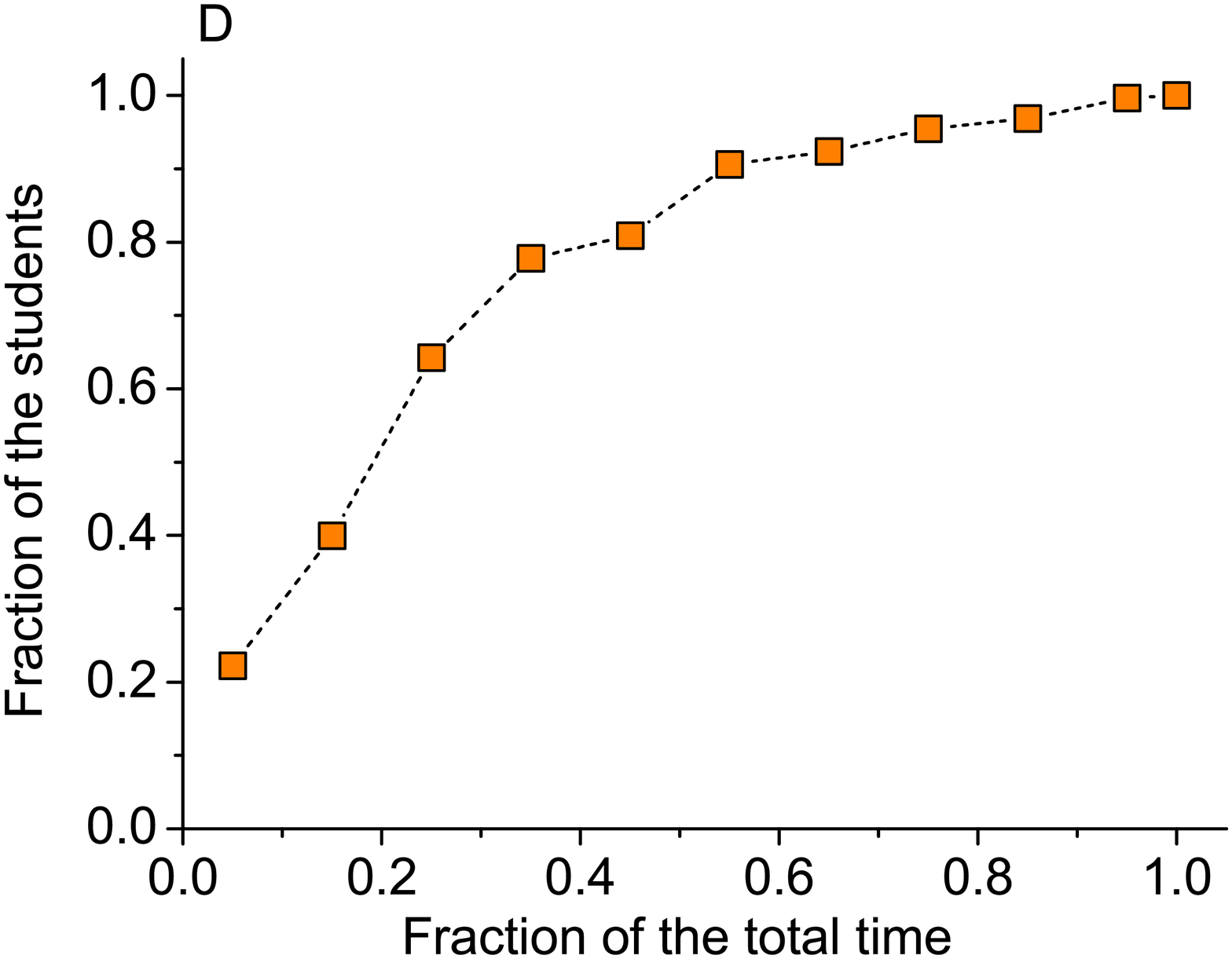}
   } }
\caption{{\bf Characterization of the dynamic network.} (A) Evolution
  of the degree distribution of the giant component of the Romantic
  network, when the frames are taken separately. (B) Degree
  distribution at three selected times ($t=20, 30, 35$) compared with
  the static degree distribution.  (C) Average degree $\left< k \right
  >_{t}$ and largest connected component size $G/N$ (normalized by
  $N=288$) as a function of time.  (D) Cumulative distribution of
  interaction times, as a fraction of the total time.
}\label{fig:dynamicnetwork}
\end{figure}

The Fig.~\ref{fig:dynamicnetwork}A displays the degree distribution at
different times. The mere inspection of that figure makes clear that
any measure we can think of for networks will give very different
values if we applied to an individual frame, instead of the static
network that results from the superposition of all the frames.  That
can be appreciated in Fig.~\ref{fig:dynamicnetwork}B, where we compare
the degree distribution at three different times around the most
active frame, with the static network degree distribution. As time
goes by, we have: for $t<15$ the interactions are only pair-like, with
no connection between different pairs at the same time; the time
window $30 < t < 40$ is where the activity increases to its maximum at
$t=35$; for $t>40$ the degree distribution returns to low values. This
is an extremely important point, because unlike some proposed models
for the spread of epidemics in dynamic
networks~\cite{Altmann,Volz2007,Rissau2010}, this scenario can not be
represented by an asymptotic degree distribution.

Greater is the effect over the average degree, $\left< k \right>_{t}$,
which has a peak value $\left< k \right>=1$ at $t \approx 35$, but
most of the time it is much lower than the static value $\left< k
\right>=2$ (Fig.~\ref{fig:dynamicnetwork}C).  Probably the most
important effect of such measure ---in fact the reason of that
difference--- is that in all frames, many individuals have no contacts
at all ($k=0$).  Following the behavior of random
networks~\cite{Erdos1959}, we observe that the variation of the
average degree directly influences the giant component size, $G$.

Another important measurement, related with the dynamic formation of
links, is the fraction of time that each individual participates in
the sexual activity, which is represented in
Fig.~\ref{fig:dynamicnetwork}D. That figure shows, for example, that
while the majority of the students ($\sim$ 90\%) participates less
than 45\% of the time, a small fraction of the them ($\sim$ 10\%) is
active during more than 55\% of the total time.

\subsubsection*{Dynamic disease}
\begin{figure}[htb!] \centerline{
\subfigure{\includegraphics[width=6cm]{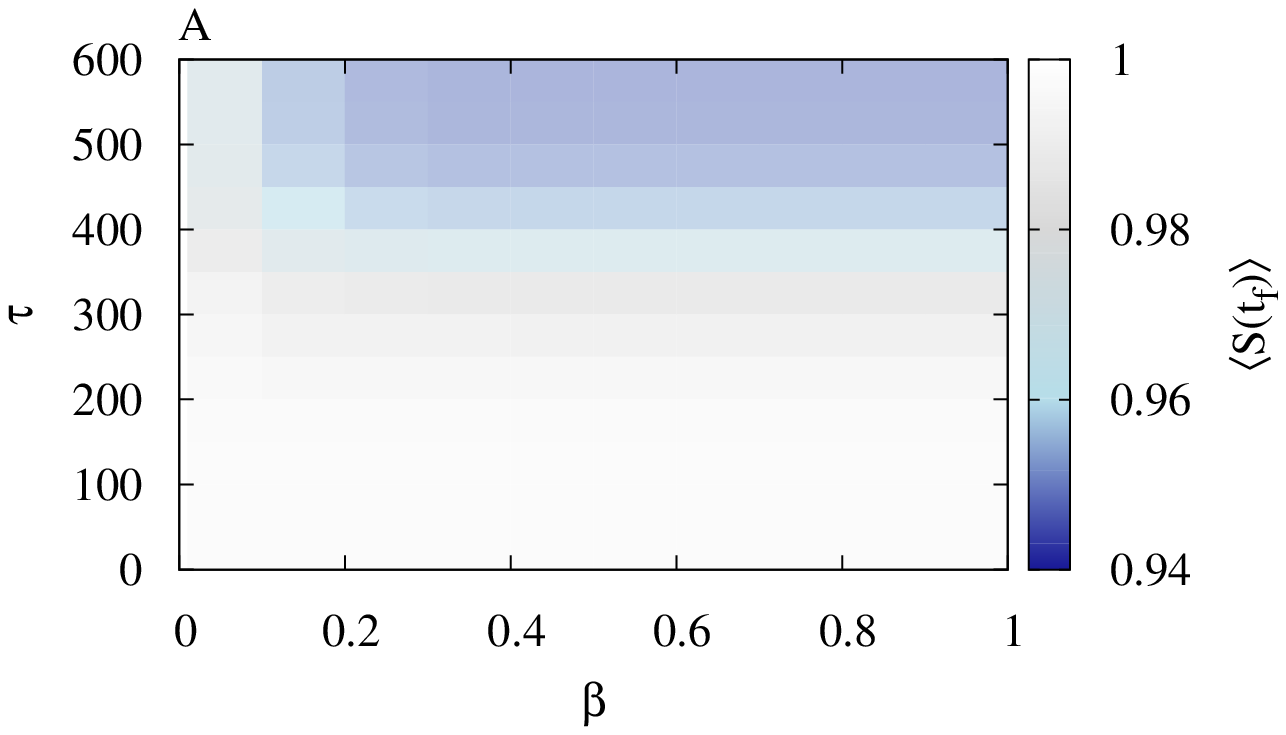}
  } \hfil
\subfigure{\includegraphics[width=6cm]{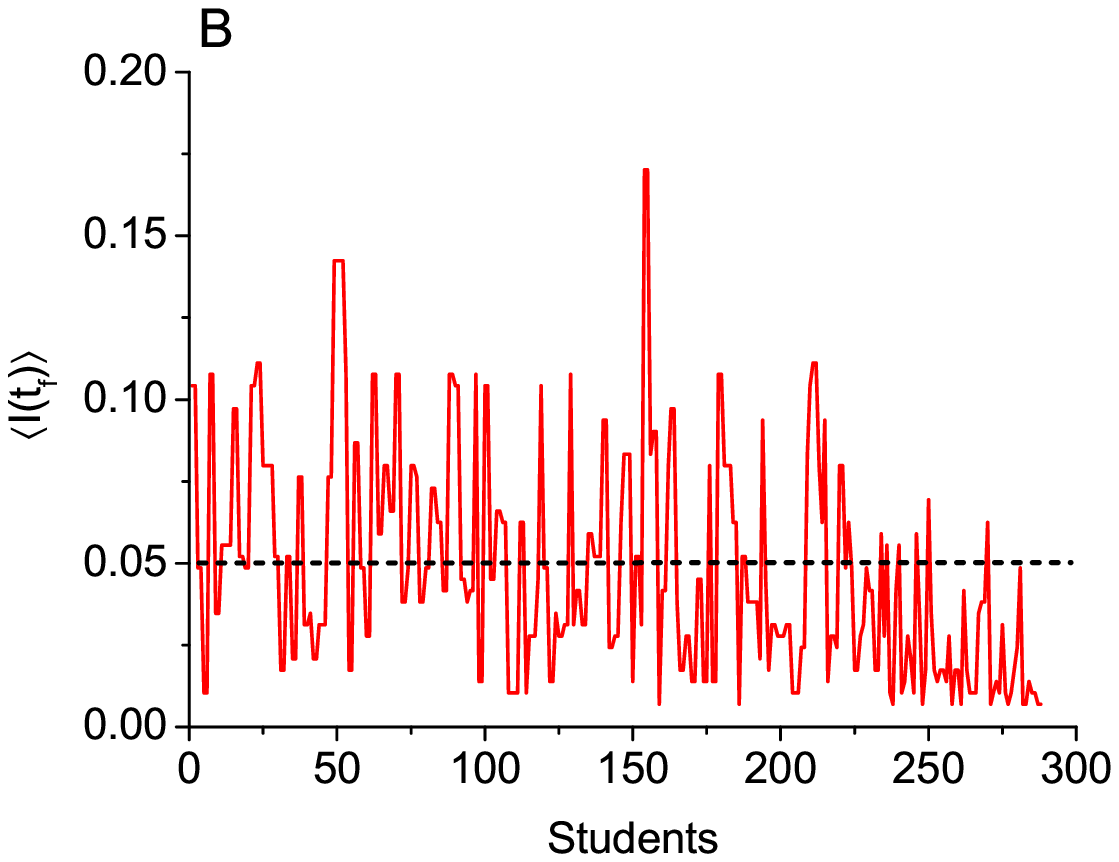}
   }}
   \centerline{
\subfigure{\includegraphics[width=6cm]{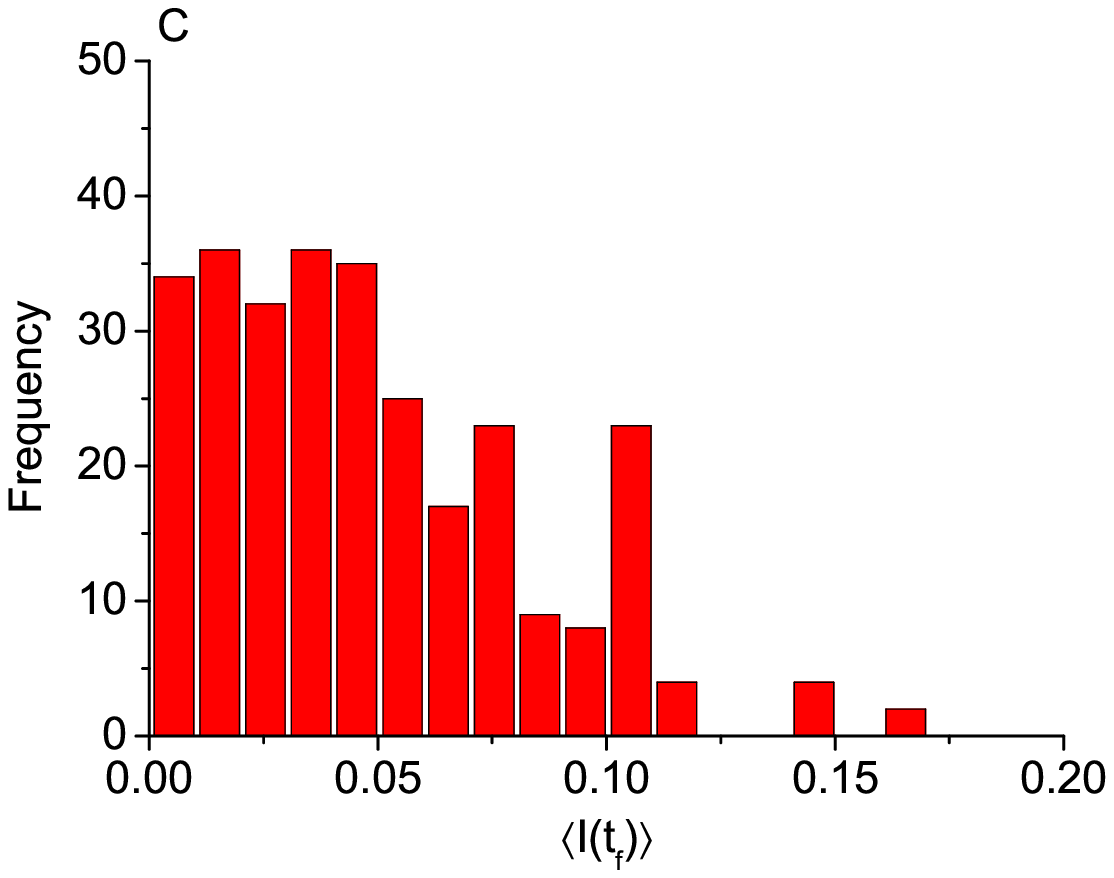}
  } \hfil
\subfigure{\includegraphics[width=6cm]{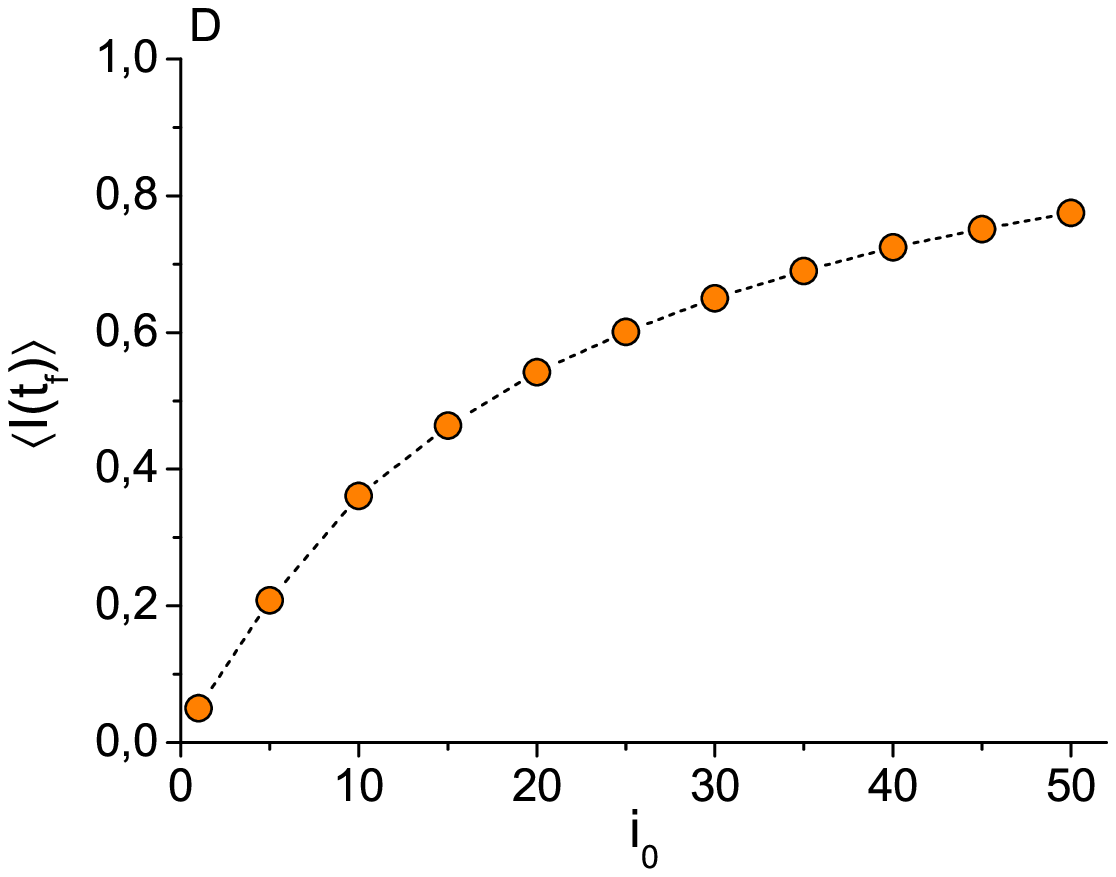}
   } }
   \centerline{
\subfigure{\includegraphics[width=6cm]{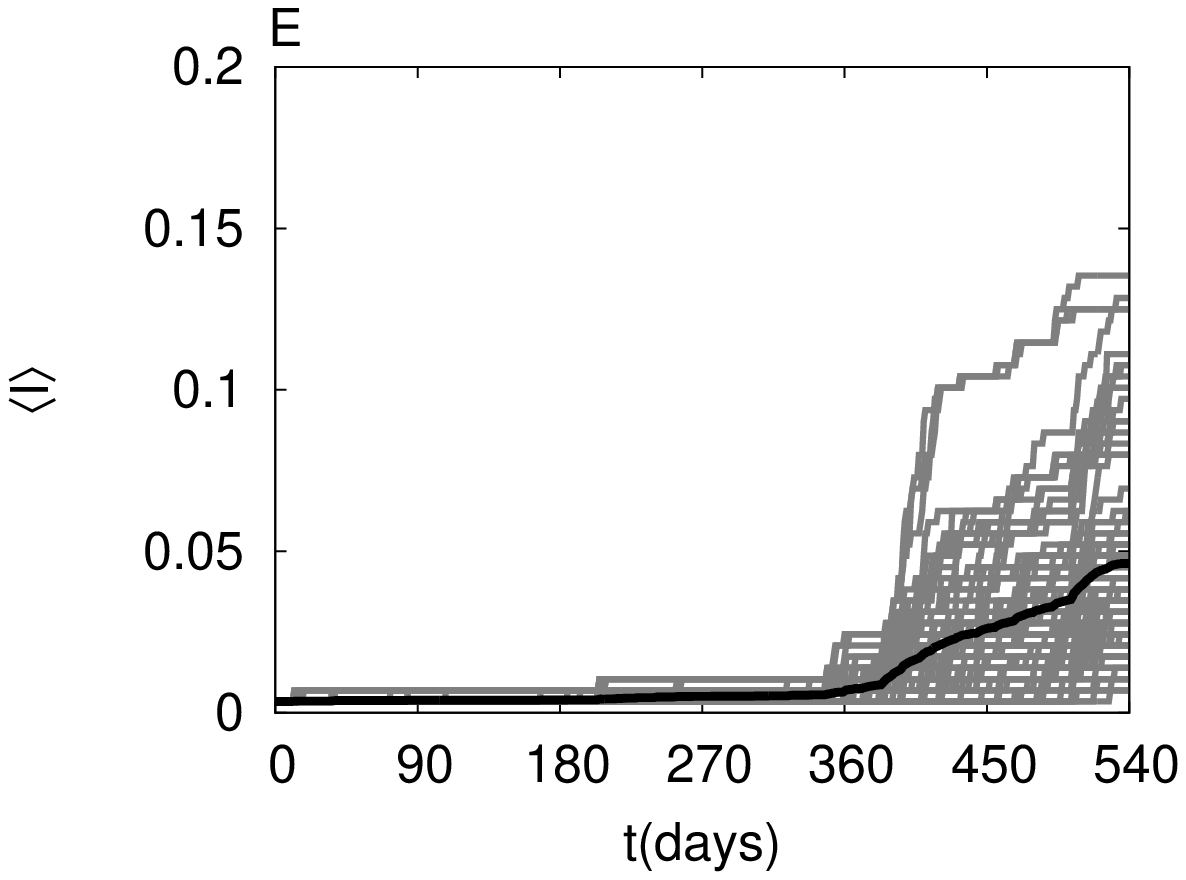}
  } \hfil
\subfigure{\includegraphics[width=6cm]{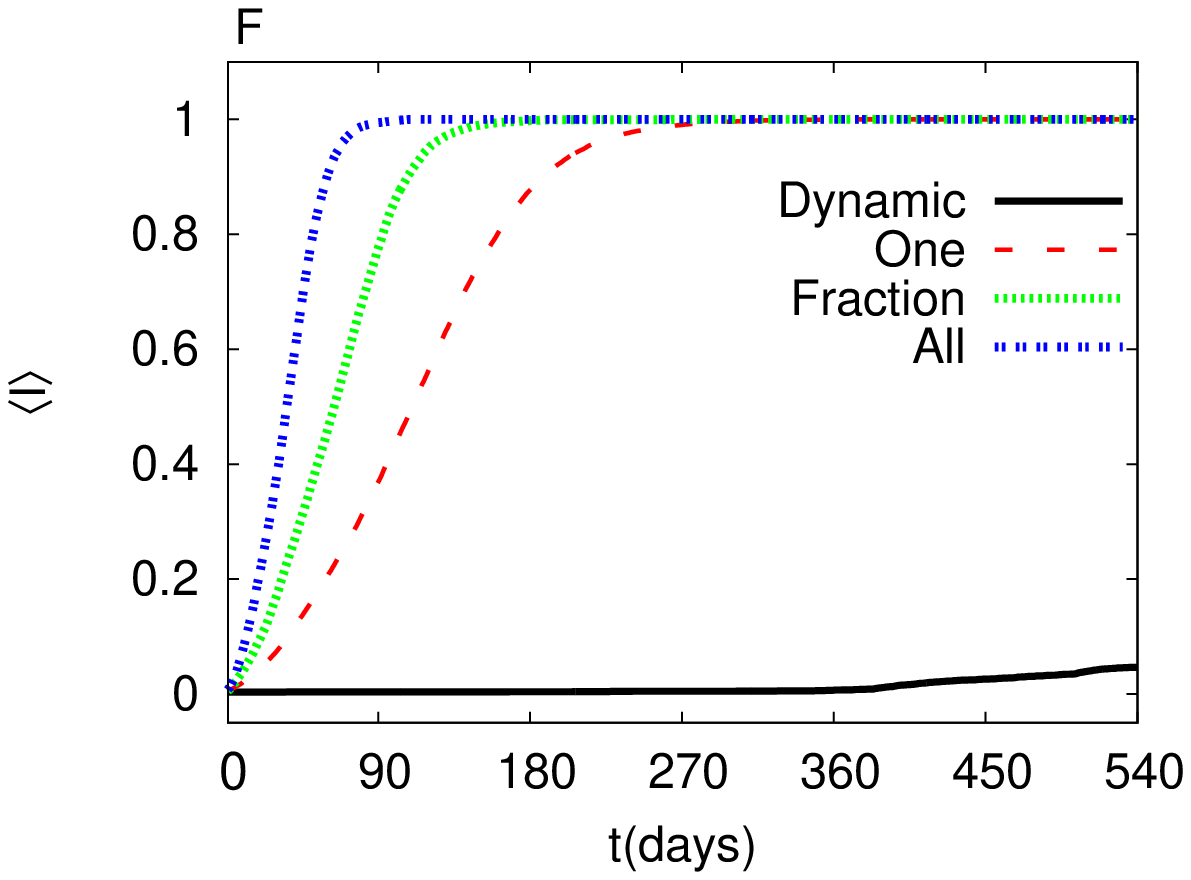}
   } }
\caption{{\bf Epidemics simulations on the ``Dynamic
Romantic network'':} (A) Final fraction of non-infected subjects
$\left<S(t_f)\right>$ as a function of the infective time $\tau$ and
the infection probability $\beta$ (white region means almost no
infection); (B) distribution of infection size (fraction of total ever
infected) depending on the initial infected subject; (C) histogram of
the distribution (B); (D) average infection size (fraction of total
ever infected) vs the number of initial infected subjects; (E)
fraction of infected subjects vs time; different runs with different
seeds in gray, average in black; (F) average number of infected
subjects vs time, comparison between dynamic network and static
network with different dynamics rules.  Each point of the curve is the
result of an average over 200 independent runs.
In figures B-F $\tau = 540$ and $\beta=0.2$ (notice however that
figure A shows that for such value of $\tau$ the dependence on
$\beta$ is weak above $0.2$).
}
\label{fig:SIRDynamic}
\end{figure}

The first result (see Fig.~\ref{fig:SIRDynamic}A,F) is remarkable: for
the wide interval of $\beta \tau$ values explored, the final size of
the epidemic is always much smaller than the corresponding static
results\footnote{For the comparison to be possible we use the same
  unit of time (a day) for the time step in both cases; each of the 45
  frames in the dynamic scenario lasts 12 days, which gives a total of
  540 days. So, in the static scenario we use a total of 540 time
  steps.}.  Evidently, no epidemic should be observed in this
scenario. In fact, we can see that only for infection times $\tau$
close to the whole duration of the dynamics, there are some effects
over the susceptible population, but very weak indeed.  Therefore, it
is interesting to consider an extreme epidemic case, the SI epidemic
dynamics, which means a lifetime infection time\footnote{Some
  diseases, notoriously infection by HPV virus, have this kind of
  dynamics where $\tau$ is equal to the lifetime.}, so the relevant
time is the duration of the dynamics. This is what is presented in
Figures \ref{fig:SIRDynamic} with parameters $\tau = 540$ and $\beta =
0.2$, except in Fig.~\ref{fig:SIRDynamic}A where all possibles values
of these two parameters are explored (up to the extreme value $\beta =
1$, which means a certain infection every time an S-I contact is
tried). Even with this extreme setup the total number of infected
individuals at the end is under 15\%. The average number of infected
however, considered all the possible students as the initial infected
one is not bigger than 5\% (see Fig.~\ref{fig:SIRDynamic}B
and~\ref{fig:SIRDynamic}C ). Just to take the situation to the very
extreme, Fig.~\ref{fig:SIRDynamic}D shows the number of final infected
as a function of the number of initially infected, and we see that
with more that 50 over 288 initial infected (17\%) the final
proportion of infected is below 80\%.  All together these results
strength the key role of the dynamic formation of pairs which bring a
different scenario for modeling epidemics in networks.

Besides, in the dynamic network, the total number of infected
individual is strongly affected by the initial condition, as can be
seen in Fig.~\ref{fig:SIRDynamic}E. In this case, who is the initial
infected is very important, not because of its personal activity, but
due to the particular sequence in which the individuals connect to
each other. It is the whole sequence which matters, not the particular
initial subject. From the dynamics we see that the number of infected
subjects grows substantially from $t>360$.

Figure~\ref{fig:SIRDynamic}F shows the comparison between the static
and dynamic Romantic network within the SI disease.  The difference is
remarkable. The static network, in any of the transmission rules, is
much more efficient for the disease propagation than the dynamic
version.

\section*{Discussion}
In this contribution we addressed the problem of STI epidemics in a
real network (the so called ``Romantic or Jefferson School network'')
in two aspects.  First, to compare epidemics in this network with
epidemics in model networks to see whether the later ones (or which
one of them) may represent a real network.  Particularly, if the
disease propagation is sensitive enough to details in the topology.
The second aspect, the effect of the dynamic formation of links and
how this can affect the epidemic dynamics, is far more relevant,
specially because of the conclusion we have drawn.

Regarding the static network and the comparison between the Romantic
network and the paradigmatic ones, the maximum incidence and the final
size or prevalence of the epidemic are sensitive to the topology for
relative large values of $\beta \tau$ ($>5$).  For smaller values of
$\beta \tau$ ($<4$), the different topologies ---except the high
heterogeneous ones--- are hard to distinguish in terms of epidemic
outcome.  Even more if we consider the wide dispersion of the overall
results due to the small size of the networks (bars in
Fig.\ref{ws},\ref{kr}).

Homogeneous networks, like the Watts-Strogatz with $p \approx 1$ or
Krapvisky-Redner with $r \approx 0$ are statistically similar between
them and with the Romantic network in terms of epidemic outcome.  This
is valid when $\beta \tau$ is not very large and it is due to the
peculiar average value of the degree ($\left< k \right>=2$) which
makes all the structures ring or tree like, as can be seen by direct
eye inspection of Fig.~\ref{fig:networks} (A-rightmost, B, and
C-leftmost).  The main difference between these artificial
realizations and the real one is its ring. However is an only and
relative large ring.  Among all artificial static networks tested in
this contribution, we can conclude that the Watt-Strogatz network with
$p \approx 1$ (random network with $\left< k \right>=2$) is
undistinguished from the Romantic network in terms of disease
spreading, for values of $\beta \tau$ up to $8$.

A marginal conclusion is that any network structure, in which not
everybody is potentially connected with anybody, including the K-R
heterogeneous networks, give a less drastic scenario than the mean
field all mixed result. The only exception is the K-R network with $r
\approx 1$ (and for $\beta\tau<1.5$) because it approaches a
scale-free network.

Considering the external field, degree of the initial infected, and
rules of interactions, our results suggest that the epidemic dynamics
is affected by all these variants.

Regarding the external field, the effect that individuals contacting
infected external subjects has is to make the network closer to a full
connected one.

Concerning the degree of initial infected (or the sexual activity of
the starter of the infection) its importance is confirmed in the
Romantic network.  At $\beta\tau=4$ for example, the chances of a
burst involving more that 5\% of the subjects triples when going from
the less to the most connected as the initial infected one. The higher
the degree of the initial infected the shorter the time to reach the
peak of the infection.

Our results on contacts frequency (rules of interactions) confirm that
the spread of an epidemic not only depends on the available number of
contacts, but also on the frequency of contacts with
them~\cite{Nordvik}, and as the frequency increases the results
approach the mean field results from below.

Going to the dynamic network situation, that is considering the
sequence of connections and disconnection between the subjects, we
arrive at very interesting conclusions.  We have shown with
simulations (the only possible tool in this case) that the dynamic
network will always cause less severe epidemics.  Because not only
there are periods in which we only see small isolated components (at
those times the infection can not jump from one component to the
others), but the time windows in which they are connected (and not all
at the same time) are relatively short. In other words, relationship
timing directly influences the course of the
epidemics~\cite{Moody2002}.

Clearly, the propagation of a disease that depends on the links, which
are created but for definite time intervals, will be strongly affected
by such dynamical process.  We have seen that the average degree in
the dynamic case has a peak of value $\left< k \right>=1$, but usually
is much lower than the static value $\left< k \right>=2$. Besides, in
the dynamic case the distribution of interaction times, absent in the
static case, is crucial for disease propagation.

By testing an hypothetical extreme epidemic situation, the
Susceptible-Infective model with maximum infection probability, we
obtained a final number of infected around 15\% in the worst case
(starting the infection with the most connected subject). However, the
average number of infected subjects, over all the possible choices of
the initial infected student is not bigger than 5\%.  Putting
together, these results make apparent the key role of the dynamic
formation of pairs, which brings a different scenario from previous
results in static networks.  Remarkably, in the dynamic network the
total number of infected individuals is strongly affected by the
initial condition. In this case, who is the initial infected is
important, not due to its personal activity, but because of the
particular sequence in which the individuals connect between them. It
is the whole sequence which matters, not the particular subject.

Moreover, comparing figures Fig.~\ref{fig:dynamicnetwork}C and
Fig.~\ref{fig:SIRDynamic}E we see a clear correlation between them.
Thus, the dynamic of the largest component is related with the
evolution of the fraction of infected subjects. Like a random network,
the giant component follows a close relation with the average degree,
as can be seen comparing $\left<k\right>_{t}$ and $G/N$ in
Fig.~\ref{fig:dynamicnetwork}C.  This leads us to conclude that the
fraction of infected people is related to the dynamics of the average
degree, which can be important for public health.

Additionally, we observe that the static network, in any of the
transmission rules, is much more efficient for the disease propagation
than the dynamic version.  The effect of the dynamic formation (and
break) of links is quite clear: A disease that depends on the
structure of the links which are not available all the time, but in
very specific time windows, will be influenced by them; but in a
``desirable way'', because the worst scenario is always the static
one, when all links are available all the time.

\section*{Acknowledgments}
We acknowledge the thorough analysis of our work and critical feedback 
from one anonymous reviewer. This research was supported from the Brazilian
agencies CNPq and CAPES. Partial support from the following projects:
CNPq-PROSUL \# 490440/2007 and join PRONEX initiative CNPq/FAPERGS \#
10/0008 is also acknowledged.


\begin{thebibliography}{10}
\bibitem{Bernoulli}
Bernoulli, D.,
\newblock Essai d’une nouvelle analyse de la mortalité causé par la
petite vérole, et des avantages de l’inoculation pour la
prévenir. Histoire de l’Acad. Roy. Sci. (Paris) avec Mém. des Math. et
Phys.  and Mém., (1760) 1. Rev. Med. Virol. {\bf 14} (2004) 275.

\bibitem{Hamer}
Hamer, W. H.,
\newblock The Lancet {\bf 167} (1906) 569.

\bibitem{Kermack1991}
Kermack, W. and McKendrick, A.,
\newblock Bulletin of Mathematical Biology {\bf 53} (1991) 33.

\bibitem{Bailey}
Bailey N.T.J.,
\newblock The Mathematical Theory of Infectious Diseases. Griffin,
London, second edition, 1975.

\bibitem{Murray}
Murray J.D.,
\newblock Mathematical Biology, Springer-Verlag, New York (2002).

\bibitem{Colizza2009}
Balcana D., Colizza V., Gonçalves B., Hud H., Ramascob J.J, Vespignani A.,
\newblock PNAS {\bf 106} (2009) 21484.

\bibitem{Colizza2011}
Bajardi P., Poletto C., Ramasco J.J., Tizzoni M., Colizza V.,
Vespignani A.,
\newblock PloS One {\bf 6} (2011) e16591.

\bibitem{Kuperman2001}
Kuperman, M. and Abramson, G.,
\newblock Phys. Rev. Lett. {\bf 86} (2001) 13.

\bibitem{Albert2002}
Albert, R. and Barabasi, A.,
\newblock Rev. Mod. Phys. {\bf 74} (2002) 47.

\bibitem{Jones2003}
Jones, J. H. and Handcock, M. S.,
\newblock  Proc. R. Soc. B {\bf 270} (2003) 1123.

\bibitem{Pastor2001}
Pastor-Satorras, R. and Vespignani, A.,
\newblock Phys. Rev. Lett. {\bf 86} (2001) 3200.

\bibitem{May2001}
May, R.~M. and Lloyd, A.~L.,
\newblock Physical Review E {\bf 64} (2001) 066112.

\bibitem{Newman2002b}
Newman, M.~E.~J., Jensen, I., and Ziff, R.~M.,
\newblock Physical Review E {\bf 65} (2002) 021904.

\bibitem{Funk2010}
Funk, S., Salathé M., and Jansen, V.~A.~A.,
\newblock J. R. Soc. Interface {\bf 7} (2010) 1247.

\bibitem{Eubank2004}
Eubank, S. et~al.,
\newblock Nature {\bf 429} (2004) 180.

\bibitem{Liljeros2001}
Liljeros, F., Edling, C.~R., Amaral, L. A.~N., Stanley, H.~E., and Aberg, Y.,
\newblock Nature {\bf 411} (2001) 907.

\bibitem{Foxman2006}
Foxman, B., Newman, M., Percha, B., Holmes, K. K. and Aral, S. O.,
\newblock Sex. Transm. Dis. {\bf 33} (2006) 209.

\bibitem{Bearman2004}
Bearman, P.~S., Moody, J., and Stovel, K.,
\newblock American Journal of Sociology {\bf 110} (2004) 44.

\bibitem{Moody-web}
Moody, J. (2012, August 06). Jefferson Romantic Ties.\\ 
\newblock Retrieved {\it http:$//$www.soc.duke.edu$/\sim$jmoody77$/$NetMovies$/$rom\_flip.html}

\bibitem{Helleringer2007}
Helleringer, S. and Kohler, H.,
\newblock AIDS {\bf 63} (2007) 2323.

\bibitem{Rocha2010}
Rocha, L.E.C., Liljeros, F. and Holme, P.,
\newblock PNAS {\bf 107} (2010) 5706.

\bibitem{Doherty2011}
Doherty I. A.,
\newblock Current Opinion in Infectious Diseases {\bf 24} (2011) 70. 

\bibitem{Morris1997}
Morris, M. and Kretzschmar, M.,
\newblock AIDS {\bf 11} (1997) 641.

\bibitem{Moody2002}
Moody, J.
\newblock Soc. Forces {\bf 81} (2002) 25.

\bibitem{Volz2007}
Volz, E. and Meyers, L.A.,
\newblock Proc. R. Soc. B {\bf 274} (2007) 2925.

\bibitem{Rissau2010}
Risau-Gusman, S.,
\newblock J. R. Soc. Interface {\bf 9} (2012) 1363.

\bibitem{Rocha2011}
Rocha, L.E.C., Liljeros, F. and Holme, P.,
\newblock PLoS Comput. Biol. {\bf 7} (2011) e1001109.

\bibitem{Batagelj1998}
Batagelj, V. and Mrvar, A.,
\newblock Connections {\bf 21} (1998) 47.

\bibitem{Watts1998}
Watts, D. J. and Strogatz, S. H.,
\newblock  Nature {\bf 393} (1998) 440.

\bibitem{Krapivsky2001}
Krapivsky, P.~L. and Redner, S.,
\newblock Phys. Rev. E {\bf 63} (2001) 0066123.

\bibitem{Marsaglia}
Marsaglia, G., Zaman, A. and Tsang, W.,
\newblock Stat. Prob. Lett {\bf 9} (1990) 35.

\bibitem{Altmann}
Altmann, M.,
\newblock J. Math. Biol. {\bf 33} (1995) 661.

\bibitem{Erdos1959}
Erdos, P. and Renyi, A.,
\newblock Publ. Math, Decebren {\bf 6} (1959) 290.

\bibitem{Nordvik}
Nordvik, M.K. and Liljeros, F.,
\newblock Sex. Transm. Dis. {\bf 33} (2006) 342.






\end{thebibliography}

\section*{Figure Legends}

\end{document}